# PHYSICS BREAKTHROUGH DISPROVES FUNDAMENTAL ASSUMPTIONS OF THE CHICAGO SCHOOL

*Cortelyou C. Kenney*[*]

## ABSTRACT

*Classical law and economics is foundational to the American legal system. Centered at the University of Chicago, with a reach touching nearly every law faculty and economics department in the country, its assumptions, most especially that humans act both rationally and selfishly, informs the thinking of legislatures, judges, and government lawyers, and has shaped nearly every aspect of the way commercial transactions are conducted. Its influence extends from the most transparently commercial enterprises and the how they are regulated to the tort law system and even to criminal law. But what if the Chicago School, as I refer to this line of thinking, is wrong? Many thoughtful scholars have argued that the Chicago School's assumptions that humans act rationally is incorrect. Alternative approaches such as behavioral law and economics or law and political economy contend that human decision-making is based on emotions or should not be regulated as a "social geometry of bargains."[1]*

*This Article proposes a different and wholly novel reason that the Chicago School is wrong: a fundamental assumption central to many of its game theory models has been disproven. More specifically, this Article shows that a 2012 breakthrough from world famous physicist Freeman Dyson "shocked the world of game theory."[2] This game theory breakthrough is now accepted in the fields of*

---

[*] Academic Fellow, Cornell Law School; Affiliated Fellow, Yale Law School; Information Society Project; UC Berkeley Concurrent Enrollment Student. For their helpful input, the author thanks Yonathan Arbel, Roxanna Altholz, Zohra Ahmed, Amanda Bashi, Kaushik Basu (Professor of Economics at Cornell University), Briana Beltran, Jeffrey Bowen (Associate Professor of Physics at Bucknell College), Alexandra Cirone (Assistant Professor of Political Science at Cornell University), Mike Dorf, George El-Khoury, Douglas Edlin, Anne Emerick, Darius Fullmer, James Grimmelman, Nik Guggenberger, Angela Harris, Rob Hendricks, Aziz Huq, Sheri Lynn Johnson, Amy Kapczynski, Bonnie Kaplan, Doug Karpa, Riley Keenan, Margaret Kwoka, Anne Kenney, John Kenney (Associate Professor of Physics at East Carolina University), Ian Kysel, Joshua Macey, Chan Tov MacNamara, Brooks Miner (Associate Professor of Biology at Ithaca College), Joshua Plotkin (Walter H. and Leonore C. Annenberg Professor of the Natural Sciences at the University of Pennsylvania), Aziz Rana, JP Schnapper Casteras, Charles Seife (Professor of Journalism at NYU), Steve Shriffrin, Nelson Tebbe, Tyler Valeska, Jakki Kelley-Widmer, and Carl Williams. The author also wishes to acknowledge the research assistance of Jamie Smith, Vincent Su (Ph.D. Candidate in Physics at the University of California, Berkeley); Steven T. Marzagalli, and Cornell Law Library librarian Jacob Seyward for their excellent support. I have engaged in advocacy work with some of the group expressed herein, but all views expressed represent my personal opinions, not the views of any affiliated institution or client, if any. Any errors are my own.

[1] Jedediah Britton-Purdy, David Singh Grewal, Amy Kapczynski & K. Sabeel Rahman, *Building a Law-and-Political-Economy Framework: Beyond the Twentieth-Century Synthesis*, 129 YALE L.J. 1784, 1824, 1827 (2020).

[2] Katherine Unger Baillie, *Penn Biologists Show That Generosity Leads to Evolutionary Success*, PENN TODAY (Sept. 3, 2013), https://penntoday.upenn.edu/news/penn-biologists-show-generosity-leads-evolutionary-success.

## *GAME THEORY*

*mathematics, engineering, and evolutionary biology, but has not made it into mainstream economics, much less law and economics, and no scholar to date has connected it to the Chicago School. This Article builds on this body of work and proves that there are new answers to game theory models that scholars in the Chicago School use as the basis for rationalizing the tort law system as well as the criminal justice system. This Article shows that Chicago School game theorists are wrong on their own terms because these 2 x 2 games such as the Prisoner's Dilemma, Chicken, and Snowdrift—ostensibly based on mutual defection and corrective justice—in fact yield to an insight of pure cooperation. These new game theory solutions can be scaled to design whole institutions and systems that honor the pure cooperation insight, holding out the possibility of cracking large scale social dilemmas like the tragedy of the commons. It demonstrates that, in such systems, pure cooperation is the best answer in the right environment and in the long run. It ends by calling for a new legal field to redesign the structures based on the outdated assumptions of the Chicago School game theorists.*

INTRODUCTION

Law has always drawn on classical law and economics, with many law and economic scholars focusing on game theory. In particular, game theory has influenced corrective justice approaches to the legal system and the rules that govern it. Classical law and economics, which I shall refer to as the "Chicago School,"[3] uses game theory as "a set of tools and

---

[3] Robin I. Mordfin & Marsha Nagorsky, *Chicago and Law and Economics: A History*, THE UNIVERSITY OF CHICAGO: THE LAW SCHOOL (Oct. 11, 2011) (describing the University of Chicago as the birthplace of modern law and economics and noting the "big wave of scholarship in law and economics [that] had taken hold and had moved into subjects concerning common law efficiency, torts, contracts, and criminal law."). Judge Richard A. Posner defines "the law and economics movement" as encompassing "property rights, of corporate and other organizations, of government and politics, of education, of the family, of crime and punishment, of anthropology, of history, of information, of racial and sexual discrimination, of privacy, even of the behavior of animals-and, overlapping all these but the last, of law." Richard A. Posner, *The Law and Economics Movement*, 77 AM. ECON. REV. 1, 1 (1987). He explains the movement makes three core assumptions: (1) people act rationally, whether to marry or to form a business; (2) rules of law operate on these activities to impose prices; (3) these rules (especially judge made rules) are "best explained as efforts, whether or not conscious, to bring about efficient outcomes." *Id.* at 5. This is the working definition of the Chicago School I use to describe theorists who embrace these assumptions. Various members of the University of Chicago may, or may not, agree with these ideas. *See* Lisa Bernstein, Richard A. Epstein, Eric Posner & Randal C. Picker, *The New Chicago School: Myth or Reality?*, 5 UNIVERSITY OF CHICAGO LAW SCHOOL ROUNDTABLE 1 (1998) (debate amongst various faculty at Chicago with differing methodologies and politics, with Eric Posner and Randal Picker centered on game theory and including abolitionists such as Tracey Meares). And Judge Posner's dedication to "neoclassical economics faltered after the 2008 financial crisis. *See* Richard A. Posner, *How I Became a Keynesian*, NEW REPUBLIC (Sept. 23, 2009), https://newrepublic.com/article/69601/how-i-became-keynesian [https://perma.cc/PY5M-HRAU]." *Neil H. Buchanan & Michael C. Dorf, A Tale of Two Formalisms: How Law and Economics Mirrors Originalism and Textualism*, 106 CORNELL L. REV. 591, 599 & n. 23 (2021). The Chicago School is also not a monolith, and, as with any group, its scholars have different views. For example, some more recent Chicago School scholars take issue with prior generations, and specifically resist the punitive instinct on which large portions of the law and economics movement are founded. *See, e.g.,* Will Dobbie, Jacob Goldin & Crystal S. Yang, *The Effects of Pretrial Detention on Conviction, Future Crime, and Employment: Evidence from Randomly Assigned Judges*, 108 AM. ECON. REV. 201 (2018) (recent Chicago School voices using law and economics to critique pretrial detention to show that while it increases flight risk, it increases risk of conviction, and has negative employment-related consequences and concluding "unless there are large general

## *GAME THEORY*

a language for describing and predicting strategic behavior" that models human behavior. It encourages the legal system to "take [interdependent] and [intertwined actions] into account."[4] Scholars like Judge Richard Posner—the heart and soul of the Chicago School[5]—have used economic and mathematical arguments "implicitly" and explicitly relying on game theory to justify our systems of antitrust law, tort law,[6] and even criminal justice.[7]

Chicago School game theorists employ canonical problems such as the Prisoner's Dilemma[8] to study how cooperation works in theory in a world in which the players are rational actors.[9] Indeed, the paradox

---

deterrence of detaining individuals before trial, releasing more defendants will likely increase social welfare "); Dan M Kahan, *What's Really Wrong With Shaming Sanctions*, 84 TEX L REV 2075 (2005) (third-generation Chicago School scholar retracting an earlier article advocating shaming as a form of punishment and embracing restorative justice, at least for some crimes)  For the purposes of this paper, I focus on scholars in the Chicago School who use mathematics and game theory to rationalize punitive economic or social positions

[4] Randal C Picker, *An Introduction to Game Theory and the Law* at 2 (University of Chicago Coase-Sandor Institute for Law & Economics Working Paper No  22, 1994)

[5] *See, e.g.*, Jedediah Purdy, *State of the Debate: The Chicago Acid Bath*, AM PROSPECT (Nov 16, 2001), https://prospect org/culture/books/state-debate-chicago-acid-bath/  (describing the influence of Judge Posner on the Chicago School)

[6] DOUGLAS G BAIRD, ROBERT H GERTNER, RANDAL C PICKER, GAME THEORY AND THE LAW 47-49 (1994) ("Landes and Posner (1987) and Shavell (1987) reveal the accumulated understanding of two decades' worth of economic analysis of torts  Neither text makes overt use of game-theoretic concepts    Arguments in these books, however, do implicitly rely upon the ideas of game theory")

[7] *See* Isaac Ehrlich & Richard A Posner, *An Economic Analysis of Legal Rulemaking*, 3 J LEGAL STUD 257 (1974) (endorsing punishment in the criminal and civil justice systems using math); *see also* Richard A Posner, *An Economic Theory of the Criminal Law*, 85 COLUM L REV 1193 (1985) (using math to justify incarceration and the death penalty); Richard A Posner, *Rational Choice, Behavioral Economics, and the Law*, 50 STAN L REV 1551, 1561-64 (1998); Richard A Posner, *Retribution and Related Concepts of Punishment*, 9 J LEGAL STUD 71 (1980).

[8] For canonical works, *see generally* Picker, *supra* note 4; BAIRD, GERTNER, & PICKER, *supra* note 6, at 1, 5, 33-34, 48-49, 167, 187-91, 194-95, 201, 217, 312-33 (1994); ERIC A POSNER, LAW AND SOCIAL NORMS 14-15 (2000); Eric A Posner, Katheryn E Spier, & Adriene Vermuele, *Divide & Conquer*, 2 J LEGAL ANALYSIS 417 (2010) (discussing other answers to the Prisoner's Dilemma to model for "labor law, bankruptcy, constitutional design and the separation of powers, imperialism and race relations, international law, litigation and settlement, and antitrust law")

[9] Note that not all members of the Chicago School assume that humans are rational actors, and not all are opposed to cooperation  *See* POSNER, *supra* note 8 at 14-15  However, Professor Eric Posner states in his book "[c]omputer studies suggest that the optimal strategy might be to defect only after a pair of defections, as a way of reducing the variance caused by noise    but in any context there will be an indefinitely large set of strategies that might seem reasonable to a player  In n-person games it is necessary (1) parties either communicate among themselves (with little error) the outcomes of rounds or observe them directly, (2) parties remember a person's history as a cooperator and defector, and (3) parties adopt extreme strategies, such as defecting perpetually after a single defection    It is not clear that one should expect players to realize these strategies are appropriate; indeed, they do not appear to correspond to real world behavior  The general point is that these models provide a case for the *possibility* of cooperation (in the face of the prisoner's dilemma's case for the impossibility of cooperation) but do not guarantee that cooperation will occur or even that the likelihood of cooperation is high "

## *GAME THEORY*

embodied by the Prisoner's Dilemma is "one of the most dominant paradigms in recent theoretical works in economics, politics, and law."[10] Other 2 x 2 games—such as Chicken and Snowdrift—"provide much of the formal apparatus at work in the norm theory literature . . . as a central tool [for] stud[y] [of] the evolution of conventions and the rise of spontaneous order."[11] And despite advances in game theory from other fields—some of which have been incorporated by books and articles written by economists in the Chicago School using advanced computer models[12]—as Yale economist Ian Ayres observes "legal scholarship has remained largely ignorant of these advances."[13] As a result, analysis of the Prisoner's Dilemma and other 2 x 2 games "continue[s] to be mindlessly mired in the game theory 'technology' of the fifties."[14]

In the Prisoner's Dilemma, two co-conspirators to a crime are separately detained, isolated, and interrogated by the police. In the Dilemma, the accepted wisdom it that is in the individual best interest of each player to "defect" and rat out the other player in exchange for immunity from prosecution.[15] The problem is that if they both rat the other out, then neither is entitled to immunity, and they both go to jail. In this reading of the Dilemma, the ideal outcome for the collective (as distinct from the individual) is to cooperate and stay silent, in which case they will receive a substantially reduced sentence. The paradox of the Prisoner's Dilemma, according to the Chicago School, is that even though the intuitive answer is to cooperate, the strictly dominant (or individual-interest-maximizing) strategy leads both players to defect, resulting in the worst possible outcome for each.[16] One version of the payoff matrix is

---

[10] Richard H. McAdams, *Beyond the Prisoner's Dilemma: Coordination, Game Theory, and Law*, 2 S. CALIF. L. REV. 209, 210-211 & n.1, 214 (2009) (there are over 3000 law review articles dedicated to the Prisoners' Dilemma even though certain scholars in law and economics have used other more sophisticated models apart from the Prisoners' Dilemma such as coordination games)

[11] Randal C. Picker, *Simple Games in a Complex World: A Generative Approach to the Adoption of Norms*, 64 U. Chi. L. Rev. 1225, 1231 (1997)

[12] *See* GAME THEORY AND THE LAW (Ed. Eric B. Rasmussen, 2008)

[13] Ian Ayres, *Playing Games with the Law,* 42 STAN. L. REV. 1291, 1294-95 (1990) reviewing ERIC RASMUSEN, GAMES AND INFORMATION: AN INTRODUCTION TO GAME THEORY 352 (1989))

[14] *Id.*; Picker, *supra* note 11, at 1248-1281 (relying on computer models drawing from evolutionary biology, though not ZD strategies); Donald Braman, Dan M. Kahan, & James Grimmelmann, *Modeling Facts, Culture, and Cognition in the Gun Debate*, 18 SOCIAL JUSTICE RESEARCH 3 (2005) (using computer models to model gun control)

[15] Picker, *supra* note 4, at 5

[16] Picker, *supra* note 4, at 4-5. Note that many scholars of the Prisoner's Dilemma have mistakenly characterized the payoff matrix as one where silence is mutually beneficial—probably because it seems so intuitive. *See generally* Page v. United States, 884 F.2d 300 (7th Cir. 1989) (Easterbrook, J.) (in a case where the panel also included Judge Posner characterizing the problem as "Two prisoners, unable to confer with one and other, must decide whether to take the prosecutor's offer: confess, inculpate the

*GAME THEORY*

displayed below and shows the four possible outcomes to the Prisoner's Dilemma.

### The Prisoner's Dilemma

|  |  | Column Player | |
|---|---|---|---|
|  |  | *Cooperate* | *Defect* |
| **Row Player** | Cooperate | R=3, R=3<br>Reward for mutual cooperation | S=0, T=5<br>Sucker's payoff, and temptation to defect |
|  | Defect | T=5, S=0<br>Temptation to defect and sucker's payoff | P=1, P=1<br>Punishment for mutual defection |

*Figure 1: Payoff matrix for the Prisoner's Dilemma (PD). Two players, row and column, can each choose to Cooperate (C) or Defect (D) in the PD game. For each of the possible game actions, each player receives a reward according to the matrix, with the row player's reward listed first. Each player acting in her self interest is better off Defecting in a single-shot game. Figure reproduced from* ROBERT AXELROD, THE EVOLUTION OF COOPERATION 19 (1984).

    The Prisoner's Dilemma has several built-in assumptions: a lack of communication between the prisoners, an inability to trust in the good faith of other players, and is predicated on the assumption that the two co-conspirators do not know each other well and thus lack a mental model of the other player or any long-term sense of obligation to them because there is no possibility of another encounter.[17] It also assumed by most theorists that the players act selfishly, and that there are no reputational consequences "that might arise from being known as a snitch or fear of reprisal for confessing."[18] The Prisoner's Dilemma has thus been

---

other, and serve a year in jail, or keep silent and serve five years  If the prisoners could make a (binding) bargain with each other, they would keep silent and both would go free  But they can't communicate, and each fears the other will talk  So both confess  Studying the Prisoner's Dilemma has led to many insights about strategic interactions  See Thomas C  Schelling, *The Strategy of Conflict*, 53-80, 119-61 (1960; 1980 rev ); Robert Axelrod, *The Evolution of Cooperation* (1984) ")

[17] Alexander J  Stewart & Joshua B  Plotkin, *Extortion and Cooperation in the Prisoner's Dilemma*, 26 PROC NAT'L ACAD SCI U S AM (PNAS) 10134 (2012)  This Article has been reviewed by Professor Plotkin and I thank him for helpful comments in translating and popularizing his research into a new legal framework

[18] Picker, *supra* note 4, at 4

## *GAME THEORY*

taken to "offer a grim," almost Hobbesian view of "social interactions" given the perceived mutual incentive to defect.[19]

But the Prisoners' Dilemma is not always so harsh. The so-called folk theorem suggests that if the game is played repeatedly with repeat player encounters, the incentives to defect diminish and cooperation is easier to sustain.[20] Study of the Iterated Prisoner's Dilemma—where players encounter one and other in a round robin—also yields a more hopeful answer. Robert Axelrod, the father of the most famous modern brand of game theory, wrote a pioneering article,[21] followed by a book,[22] and subsequent researchers building on his work generated tens of thousands of articles on the solution he identified. On Robert Axelrod's view, the "winning" strategy in Iterated Prisoner's Dilemma is referred to as "Tit for Tat," which very much resembles the Biblical thinking of an "Eye for an Eye"[23] to right a wrong and to deter future misconduct.[24] "Tit for Tat" is premised on the principle of "reciprocity" and "represents a balance between punishing and being forgiving."[25] "Tit for Tat" models are loosely based the notion that good behavior should be rewarded in kind, and bad behavior should be punished, though one should start with an assumption of good behavior and be forgiving assuming an individual has reformed.[26]

---

[19] Stewart & Joshua B Plotkin, *supra* note 17, at 10134
[20] Lones Smith, *Folk Theorems in Overlapping Generations Games*, 4 GAMES & ECON BEHAV 426 (1992)
[21] Robert Axelrod, *Effective Choice in the Prisoner's Dilemma*, 24 J CONFLICT RESOLUTION 1, 3 (1980)
[22] ROBERT AXELROD, THE EVOLUTION OF COOPERATION 19 (1984) E O Wilson also tremendously influenced this field, authoring the books such as *Sociobiology: The New Synthesis* (1975) that also focused on cooperation and questions such as reciprocal altruism, though not the Prisoner's Dilemma
[23] *Id.* at 173
[24] This maxim is enshrined in numerous places within the Bible, primarily in the Old Testament In Leviticus, it is said: "Whomever takes a human life shall surely be put to death Whoever takes an animal's life shall make it good, life for life If anyone injures his neighbor, as he has done it shall be done to him, fracture for fracture, eye for eye, tooth for tooth; whatever injury he has given a person shall be given to him " *Leviticus 24:18-20* But the New Testament takes a different view: "You have heard that it was said, 'An eye for an eye and a tooth for a tooth ' But I say to you, Do not resist the one who is evil But if anyone slaps you on the right cheek, turn to him the other also And if anyone would sue you and take your tunic, let him have your cloak as well And if anyone forces you to go one mile, go with him two miles Give to the one who begs from you, and do not refuse the one who would borrow from you " *Matthew 5:38-42 See also* Posner, *Retribution and Related Concepts of Punishment*, supra note 7 at 79 (arguing "the vengeful component in our genetic makeup remains an important element in deterring aggression today Nuclear deterrence is premised on the belief that a nation's leaders will retaliate in circumstances (the complete destruction of a nation) where retaliation could yield no tangible benefits Another example is the belief that people will terminate trading relations with those who have cheated them without calculating the costs and benefits of continuing those relations—without, that is, treating the cost to them of the wrong sunk cost ")
[25] AXELROD, *supra* note 22, at 45
[26] "Tit for Tat" ultimately won both tournaments run by Axelrod and cemented the view that the best way of promoting cooperation is to be forgiving, and to mete out punishment for betrayals Other

## *GAME THEORY*

The Chicago School's use of game theory has, in turn, influenced other legal fields from institutional design,[27] to negotiation,[28] to administrative law,[29] to trade law,[30] to restorative justice,[31] and to alternative dispute resolution and family law[32] and even to customary international law.[33] Other scholars have used game theory and the Chicago School's insights to suggest, for example, new approaches in antitrust law to understanding the formation of cartels,[34] as well as employer wages,[35] and the admissibility of expert evidence.[36] Game theory has been criticized in First Amendment scholarship for promoting conformism and censorship.[37] And research about cooperation in the midst of dissolving unions

---

programs performed well in the tournament, including one that Axelrod identified as potentially superior to "Tit for Tat" called DOWNING that estimated probabilities but was "doomed" because it defected on the first move  Other, more generous and forgiving rules, such as a "Tit for Two Tats," would have won the first round of the computer tournament but in the second round, where it was actually submitted, "it did not even score in the top third" because there were "some rules that were able to exploit its willingness to forgive isolated defections " Two "Tit for Tats" playing against each other would lead to an advantageous outcome, with substantially improved cooperation, but when playing against a player who defects, it can be very harsh indeed, especially if defects are repeated, as DOWNING did

[27] *See* Matthew C  Stephenson, *Information Acquisition and Institutional Design*, 124 Harv  L  Rev  1422, 1462-63 (2011); Duncan Kennedy, *Form and Substance in Private Law Adjudication*, 89 Harv  L  Rev  1685 (1976)

[28] *See, e.g.*, Robert J  Condlin, *Bargaining in the Dark: The Normative Incoherence of Lawyer Dispute Bargaining Role*, 51 Md  L  Rev  1, 104 (1992); Ronald J  Gilson & Robert H  Mnookin, *Disputing Through Agents: Cooperation and Conflict Between Lawyers in Litigation*, 94 Colum  L  Rev  509 (1994)

[29] Yehonatan Givat, *Game Theory and the Structure of Administrative Law*, 81 U  Chi  L  Rev  481 (2014); Jason Scott Johnston, *A Game Theoretic Analysis of Alternative Institutions for Regulatory Cost-Benefit Analysis*, 150 U  Pa  L  Rev  1343 (2002); Matthew C  Stephenson, *The Strategic Substitution Effect: Textual Plausibility, Procedural Formality, and Judicial Review of Agency Statutory Interpretations*, 120 Harv  L  Rev  528, 566 (Appendix) (2006)

[30] Renê Guilherme S  Medrado, *Renegotiating Remedies in the WTO: A Multilateral Approach*, 22 Wis  Int'l L J  323, 335–36 (2004)

[31] Douglas H  Yarn & Gregory Todd Jones, *A Biological Approach to Understanding Resistance to Apology, Forgiveness, and Reconciliation in Group Conflict*, 72 L  & Contemp  Probs  63, 81 (2009)

[32] Valerie A  Sanchez, *Back to the Future of Adr: Negotiating Justice and Human Needs*, 18 Ohio St  J  Disp  Resol  669, 714–15 (2003)

[33] Jack L  Goldsmith & Eric A  Posner, *A Theory of Customary International Law*,

[34] Christopher R  Leslie, *Trust, Distrust, and Antitrust*, 82 Tex  L  Rev  515, 528 (2004); Baird, Gertner, & Picker, *supra* note 6, at 166-67

[35] *See generally* Suresh Naidu, Eric A  Posner & Glen Weyl, *Antitrust Remedies for Labor Market Power*, 132 Harv  L  Rev  536, 540-547 (2018)

[36] 509 U S  579 (1993)  *See* Malcolm B  Coate & Jeffrey H  Fischer, *Daubert, Science, and Modern Game Theory: Implications for Merger Analysis*, 20 Supreme Ct  Econ  Rev  125, 126 & n 4 (2012) (citing *FTC v Swedish Match*, 131 F Supp 2d 151 (DDC 2000) and *United States v Oracle*, 331 F Supp 2d 1098 (ND Cal 2004)

[37] *See* Stephen E  Gottlieb & David Schultz, *The Empirical Basis of First Amendment Principles*, 19 J L  & Pol  145, 175 (2003)

## *GAME THEORY*

has influenced family law.[38] Preeminent thinkers have drawn on the idea of social networks to think or rethink cooperation.[39]

The Chicago School has also influenced other fields such as legislation, with prominent scholars such and William Eskridge and Philip Frickey (and their protégés) devising "public choice theory" to explain interaction between the various branches of government as "the product of a sequential game among branches, with each branch behaving strategically to enact a preferred policy" in "stylized spatial models to illustrate these interactions."[40] There are even famous cases that rely on game theory expert testimony about competition, including Whole Foods and its intended merger with Wild Oats grocery store, and Heinz and its intended merger with Beech-Nut, both of which the FTC attempted to stop, and did ultimately stop.[41]

But what if the Chicago School is wrong? Many thoughtful scholars have argued that the Chicago School's assumptions that humans act rationally is incorrect and have invented alternative approaches, most famously a body of literature focused on biology and "irrational behavior"[42]

---

[38] *See* Robert H. Mnookin & Lewis Kornhauser, *Bargaining in the Shadow of the Law: The Case of Divorce*, 88 YALE L.J. 950 (1979).

[39] *See* YOCHAI BENKLER, THE WEALTH OF NETWORKS: HOW SOCIAL PRODUCTION TRANSFORMS MARKETS AND FREEDOM (2006).

[40] William Eskridge & Philip Frickey, *The Supreme Court 1993 Term Foreword: Law As Equilibrium*, 108 HARV. L. REV. 26, 28-29 (1994); William N. Eskridge, Jr. & John Ferejohn, *The Article 1, Section 7 Game*, 80 GEO. L.J. 523, 523 (1991); William N. Eskridge, Jr., *Overriding Supreme Court Statutory Interpretation Decisions*, 101 YALE L.J. 331, 336-43 (1991); *see also* Howard S. Erlanger & Thomas W. Merrill, *Institutional Choice and Political Faith*, 22 L. & SOC. INQUIRY 959, 974–92 (1997) (citing RICHARD A. POSNER, OVERCOMING LAW (1995)); Jennifer Nou & Edward Stiglitz, *Regulatory Bundling*, 128 YALE L.J. 1174 (2019); Cass Sunstein, *Congress, Constitutional Moments, an Cost-Benefit State*, 48 STAN. L. REV. 247, 251 (1996); Laurence H. Tribe & Patrick O. Gudridge, *The Anti-Emergency Constitution*, 113 YALE L.J. 1801, 1808 (2004); Josh Benson, *The Guantánamo Game: A Public Choice Perspective on Judicial Review in Wartime*, 97 CALIF. L. REV. 1219, 1220 (2009).

[41] See Coate & Fischer, *supra*, at 175, 182 n.237; F.T.C. v. Whole Foods Mkt., Inc., 502 F. Supp. 2d 1, 17–21 (D.D.C. 2007); *rev'd*, 533 F.3d 869 (D.C. Cir. 2008), *opinion amended and superseded,* 548 F.3d 1028 (D.C. Cir. 2008). *See* 548 F.3d at 1053 (Kavanaugh, J., dissenting); *F.T.C. v. H.J. Heinz Co.*, 116 F. Supp. 2d 190, 197 (D.D.C. 2000), *rev'd,* 246 F.3d 708 (D.C. Cir. 2001).

[42] Owen D. Jones, *Time-Shifted Rationality and the Law of Law's Leverage: Behavioral Economics Meets Behavioral Biology*, 95 NW. U. L. REV. 1141, 1141–45 (2001) (describing the field born out of the influence of evolutionary biology known as "behavioral law and economics" (BLE)). This Article is completely compatible with behavioral law and economics insofar as the answers it presents to the Prisoner's Dilemma and other 2 x 2 games match human intuition and laboratory experiments. *See* Picker, *supra* note 4, at 5; KAUSHIK BASU, THE REPUBLIC OF BELIEFS: A NEW APPROACH TO LAW AND ECONOMICS 60 (2018). Basu, a professor in the Economics Department at Cornell University, explains via a problem he invented known as the Traveler's Dilemma wherein two players are asked to write down an integer between 2 and 100. If both write the same number, a third person will pay each of them that amount in dollars. If they write down different numbers, the third party will take the lower number and pay an additional reward of $2 to the person who selected it and levy a punishment to the individual who wrote the higher amount. According to Professor Basu, the Nash equilibrium to this problem is

## *GAME THEORY*

invented by Christine Jolls, Cass Sunstein, and Richard Thayer in their development of the legal theory of behavioral law and economics[43] and "choice architecture."[44] Other scholars have critiqued the emphasis on efficiency as a miseducation that they claim necessitates "a law and political economy framework."[45] They state that this attention to "political economy" requires awareness of how "economic and political power are inextricably intertwined with racialized and gendered inequity and subordination."[46] But, in their view, this law and political economy reorientation would deprioritize the "social geometry of bargains."[47] Fundamentally, their argument "requires a shift . . . of interpersonal relations—not as presumptively equal market transactions that are further legitimated by being voluntary and theoretically 'making everyone better off[,]'" but "power-laden bargains that require law and policy to be rendered more equal and fair."[48] But as Saule T. Omarova writes in her article tackling the banking system and proposing wholesale change: "Ultimately, it takes a system to beat a system"—one which must "reimagin[e] its fundamental structure and redesign[] its operation."[49]

---

for both players to agree to the lowest possible integer, in other words $2 The reason, under current game theory, is that if both players were to pick $100, a player could "deviate" and pick 99 as an integer and make $101, and so on and so forth until the best strategy is to bet the lowest But, as Professor Basu notes, there is a "large literature, experimental and theoretical, showing how the formal game theoretic prediction is not right " *Id.* His book has a chapter to explaining why lab experiments prove that real players are far more generous, typically choosing between $95 to $100, showing the assumption of "rationality"—at least as defined by the Chicago School—is incorrect and that we need a "rational rejection of rationality," *i.e.*, a world where both players recognize it is to their mutual advantage to cooperate *Id.* at 244-262 *See also* Kaushik Basu, *The Traveler's Dilemma: Paradoxes of Rationality in Game Theory*, 84 Am Econ Rev 391 (1994)

[43] Christine Jolls, Cass R Sunstein, & Richard Thaler, *A Behavioral Approach to Law and Economics*, 50 Stan L Rev 1471, 1492 (1998) (using a classic game theory game known as the "ultimatum game" to explain that "[p]eople will often behave in accordance with fairness considerations even when it is against their financial self-interest and *no one will know* ")

[44] Richard H Thaler & Cass Sunstein, Nudge: The Final Edition (2021) (categorized by Amazon as Applied Game Theory); *see also* Cass R Sunstein, *The Storrs Lectures: Behavioral Economics and Paternalism*, 122 Yale L J 1826 (2013); Nick Chater & George Loewenstein, *The i-frame and the s-frame: How focusing on individual-level solutions has led behavioral public policy astray*, Behavioral and Brain Sciences 1-60 (forthcoming draft © 2022), https://doi org/10 1017/S0140525X22002023

[45] In so doing, I answer a recent call to propound a new vision that "seek[s] to win the assent of others by offering cooperation that enables others to achieve vocation or flourishing" and in which "those who live in that order" deserve "equal weight to all members in structuring our shared life " Britton-Purdy, Grewal, Kapczynski & Rahman, *supra* note 1, at 1824, 1827

[46] *Id.*
[47] *Id.*
[48] *Id.* at 1823
[49] Saule T Omarova, *The People's Ledger: How to Democratize Money and Finance the Economy*, 74 Vand L Rev 1231 (2021)

*GAME THEORY*

This Article bucks this trend and is the first to show using what are called generous Zero Determinant (ZD) strategies that that Chicago School models on which much of the existing law and economics literature, and in turn our legal system, is rationalized are mathematically outdated.[50] It is well understood that accepted answers to Iterated Prisoner's Dilemma or other 2 x 2 repeated games are generous and forgiving to the extent an individual has not committed a wrong,[51] or has been punished for their wrong,[52] instead of deterring wrongdoing, in practice it can lead to a downward spiral, perpetuating what is known as the cycle of violence.[53] Even Robert Axelrod, the scholar who popularized "Tit for Tat" as the best model for interactions, observed that "a single defection can set off a long string of recriminations and counterrecriminations" and that should serve as a "warning against the facile belief that an eye for an eye is necessarily the best strategy."[54] And Judge Posner, who along with certain other Chicago School scholars, advocates a legal system based the concept "*lex talionis* of early Roman law, the 'eye for eye precept in the Old Testament (and a virtually identical precept in the Koran),[55] acknowledges the danger of "an endless cycle of injury, retaliation, and counter-retaliation—a costly system for controlling aggression."[56] Other answers to the Prisoner's Dilemma propounded by advocates of the Chicago School also show breakdowns in cooperation. Some are harsher, using trigger strategies, which punish even a single "defect" with a string of defects to incentivize cooperation in the

---

[50] Alexander J Stewart & Joshua B Plotkin, *Extortion and Cooperation in the Prisoner's Dilemma*, 26 PROC NAT'L ACAD SCI U S AM (PNAS) 109 (2012)  This Article has been reviewed by Professor Plotkin and I thank him for helpful comments in translating and popularizing his research into a new legal framework

[51] *See* Axelrod, *supra* note 26, describing how "Tit for Tat" operates, by cooperating on the first move and reciprocating the moves of its opponents thereafter

[52] *Id; see* AXELROD, *supra* note 26, at 71 (noting that "Tit for Tat has a memory of only one move")  But note that trigger strategies are a game theoretic answer that only punishes after the other player makes a single defection

[53] AXELROD, *supra* note 26 at 51

[54] *Id. See also* SCIENCE HISTORY PODCAST, EPISODE 17, COOPERATION: ROBERT AXELROD, https://sciencehistory libsyn com/episode-17-cooperation-interview-with-robert-axelrod (Apr 11, 2019)  22:18)

[55] Posner, *Retribution and Related Concepts of Punishment*, supra note 7, at 71 (adopting a retributivist approach to criminal law on the basis of law and economics)

[56] *Id.* at 82  But Judge Posner viewed this as a problem with "primitive" societies that responded beyond what was proportional to the crime and relied on community enforcement  *See* Richard A Posner, *A Theory of Primitive Society, With Special Reference to Law*, 23 J L & ECON 1, 5-9 (1980) (arguing "primitive" societies—characterized by their belief in "magic and sorcery," lack of "system of writing and consequently no records, and lack [of] modern communication technology were also characterized by lack of an "[]effective[] government")  He goes on to mention the problems with detecting wrongdoing, tying in the concept of deterrence  *Id.* (discussing difficulties with enforcement)

## *GAME THEORY*

"indefinitely-repeated Prisoner's Dilemma," which they themselves admit "is met by the reversion to the Nash equilibrium of (confess, confess) in the next period and in every period after that."[57]

Chicago School and law and economics scholars also rely on coordination games and embedded games incorporating the Prisoner's Dilemma. For example, Douglas G. Baird, Harry A. Bigelow Distinguished Service Professor of Law, and Robert H. Gertner, Joel F. Germunder Professor of Strategy and Finance, and Randal Picker, James Parker Hall Distinguished Service Professor of Law, all at the at the University of Chicago, discuss a version of the Prisoner's Dilemma as an embedded game where the players do not move simultaneously, and suggest playing games this way "might actually help[] the players to achieve the outcome that is in their joint interest" and provide a model for the civil legal system but "[t]hese effects, of course, could be harmful as well as beneficial."[58] Picker, in a solo-authored article, does suggest that embedded coordination games can lead to pure cooperation, stating that bringing "these two games together in a single larger game" means that "the very existence of the Prisoner's Dilemma makes it possible to coordinate on a particular Nash equilibrium in the coordination."[59] But the underlying sources he cites for this proposition suggest that "the process of iteration . . . is sufficiently involved that" that the outcome may be different than predicted because "like chains of backward induction, chains of forward induction become more suspicious as they grow longer," and scenarios where players must reason that when the other players are acting counterintuitively (in the authors words "crazy") means players "always assign a small but nonzero possibility that their opponents' playoffs being very different than originally supposed."[60] As Picker himself admits in a later paper, "[t]he Von Neumann variation suggest[s] that the basic problems of the coordination game might persist," a "disappointing" result, though one that did "suggest that the possible loss of value from inadequate coordination is naturally self-limiting."[61]

Finally, cutting-edge game theorists have more recently embraced computer simulations—including drawing on evolutionary

---

[57] Posner, Spier, & Vermuele, supra note 8, at 425
[58] *See* BAIRD, GERTNER, & PICKER, *supra* note 6, at 192-195
[59] Picker, *supra* note 4, at 19
[60] DREW FUDENBERG & JEAN TIROLE, GAME THEORY 464 (1991)
[61] Picker, *supra* note 11 at 1281-82

*GAME THEORY*

biology—to model 2 x 2 repeated games.[62] The literature they draw on "reports on the uses of computers to study self-organization in social systems. . . developments spill over from the complexity and artificial life research in biology and the physical sciences, which maintains a similar emphasis on computer simulations of complex adaptive systems."[63] No less than Robert Axelrod himself has embraced what is known as "complexity studies" that embrace a form of math known as "agent-based modeling."[64] But despite their cutting-edge nature, and the possibility of studying large systems from bacteria to armed conflict, the Chicago School has only superficially engaged with these developments, and has primarily left these developments to the work of complexity scholars elsewhere seeking to understand the legal system using agent-based modeling.[65]

Generous ZD strategies solve the cooperation breakdowns exhibited in Chicago School game theorists in two ways. First, generous ZD strategies show that intuitive answers to 2 x 2 games like the Prisoner's Dilemma—at least in an iterated game—are correct. The players should not rat each other out but should engage in a pact of pure cooperation. University of Pennsylvania researchers Alexander Stewart and Joshua Plotkin found that generous ZD strategies that fostered cooperation by using stochastic (or probabilistic) moves had an even higher payoff than "Tit for Tat."[66] In particular, the strategy that received the highest score in the tournament that they called "ZDGTFT-2," which "force[d] the relationship $S_x - P = 2(S_y - P)$ between the two players' scores"—"offering a higher portion of payoffs above P"—"received the highest total payoff, higher even that Tit for Tat and Generous-Tit-for-

---

[62] *Id.*

[63] *Id. at* 1234

[64] ROBERT ALEXROD, THE COMPLEXITY OF COOPERATION: AGENT-BASED MODELS OF COMPETITION AND COLLABORATION (1997)

[65] *See* B Ruhl, Daniel Martin Katz, & Michael J Bommarito II, *Harnessing legal complexity: Bring tools of complexity science to bear on improving law,* 355 SCIENCE 1377 (2017) ("What is often poorly recognized in these initiatives is that legal systems are also complex adaptive systems"); Matthew Koehler, *Jurisprudence Meets Physics*, 10 FRONT PHYS 760780 (2022) ("[T]he use of agent-based models has dramatically increased over the past 20 years, however they remain largely absent from the jurisprudential literature"); Joshua M Epstein, *Modeling civil violence: An agent-based computational approach*, 99 PROC NAT'L ACAD SCI U S AM (PNAS) 7743 (2002)("Agent-based methods offer a novel and, I believe, promising approach to understanding the complex dynamics of decentralized rebellion and interethnic civil violence "); EDDIE LEE, QUANTITATIVE MODELING OF COLLECTIVE BEHAVIOR (2019) (Ph D dissertation, Cornell University) (on file with author) (physics thesis modeling the U S judiciary)

[66] Stewart & Plotkin, *supra* note 17, at 10135

*GAME THEORY*

Tat, the traditional players."[67] Second, generous ZD strategies model how to create, sustain, and encourage what game theorists call prosocial behavior, or "win-win" solutions.[68] Populations can remain stable when people cooperate "only with players with good reputations," or, in the alternative, "empathy can reduce the rate of misunderstandings and unjustified defection," and "empathy can itself evolve through social contagion, inducing high rates of cooperation typical of societies that an established public monitoring system."[69]

This Article proposes the latter. It suggests the radical—though scientific—idea that there is no inherent conflict between the interest of the individual and the interest of a community, and that, in fact, they are aligned in the long run in communities that encourage empathy and connection, not exploitation.[70] While of course there may be real conflicts at the political or institutional level, ultimately this Article seeks to pose the question of institutional design and how to design, create, sustain, and scale communities that epitomize cooperation to understand the features that make them work to pave the way for a new system that embodies their values.

This Article provides the framework for a game theoretic model in two parts. Part I argues the Chicago School models the legal system as one based on corrective justice. In the tort context, the Chicago School uses game theory to rationalize a return the status quo *ex ante* by paying restitution to right wrongs. In the criminal context, the emphasis is on retribution and deterrence—namely, punishing others for the misdeeds done to us both to fulfill the instinct for revenge and to deter future misconduct. However, neither form of justice ultimately gives the

---

[67] *Id.* (citing text from Figure 1 and noting that an extortion strategy "received a lower total payoff (because its opponents are not evolving), but it won more head-to-head matches than any other strategy, save Always Defect")

[68] Alexander Ehlert, Martin Kindschi, René Algesheimer, & Heiko Rauhut, *Human social preferences cluster and spread in the field*, 117 PROC NAT'L ACAD SCI U S AM (PNAS) 22787, 22790 (2020) (noting that cooperation is contagious among friendship networks and that it can spread and turn defectors into cooperators, causing them to engage in prosocial behavior)

[69] *Id.* The authors note, however, "inferences about the perspective of another person are not always accurate, and the benefits of empathy are vulnerable to the possibility of deception and manipulation " *Id.*

[70] Even Judge Posner agrees with this statement—albeit because he believes that both individuals and the collective act selfishly See Posner, *Rational Choice, Behavioral Economics, and the Law, supra* note 7, at 1557 ("All that is required to understand altruism as a form of rational self-interest is the assumption of interdependent utilities If your welfare enters positively into my utility function, then I can increase my own welfare by increasing your welfare; and if it enters negatively, then I can increase my welfare by reducing yours ")

*GAME THEORY*

victim or survivor what they lost and neither system effectively prevents future wrongdoing.

Part II calls for a response to the Chicago School based on generous ZD strategies that are designed to promote cooperation. It explores how generous ZD strategies yield new answers to classical game theory dilemmas. And it calls for new legal field that studies these new answers with the goal of scaling them into new institutions built on atonement, compassion, and forgiveness. A particular emphasis is given to enforcement mechanisms, which have vexed scholars who study cooperation and institutional design for years. The Article concludes by suggesting law schools can one day become the types of communities these scholars envision.

I. THE CHICAGO SCHOOL AND CORRECTIVE JUSTICE

This PART explores the Chicago School's emphasis on corrective justice and game theory.[71] Even before early Chicago School scholars like Pareto and Coase focused on utility maximization,[72] classical economics was already enmeshed in game theory. Leonard Sand explains that "[a]lthough game theory initially came from outside as a critical contribution, it has now been completely embraced by the economics discipline," and that "[w]ithin economics, particular areas such as microeconomic theory, industrial organi[z]ation, international trade, and experimental economics have all been reshaped under [game] theory's influence."[73]

Sand chronicles how game theory arose—somewhat surprisingly—out of radical leftist opposition to Nazi Germany and its early focus centered on the cooperative game theory scholarship of the mathematician-refugee (also trained in mathematical physics) John Von Neumann who turned to Princeton University as a home away from Europe

---

[71] See Richard A Posner, *The Role of the Judge in the Twenty-First Century*, 86 BOSTON U L REV 1049, 1057 (2006) (discussing corrective justice and attributing it to Aristotle)

[72] See R H Coase, *The Problem of Social Cost*, 3 J LAW & ECON 1, 1-44 (1960) (advocating for a costless legal system that based on law and economics)

[73] *See* ROBERT LEONARD, VON NEUMANN, MORGENSTERN, AND THE CREATION OF GAME THEORY FROM CHESS TO SOCIAL SCIENCE, 1900–1960 (HISTORICAL PERSPECTIVES ON MODERN ECONOMICS) 1-2 (2010); *see also* Ken Binmore, *Book Review of Von Neumann, Morgenstern, and the Creation of Game Theory: From Chess to Social Science, 1900–1960*, 44 HISTORY OF POLITICAL ECONOMY 546, 546-50 (2012)

*GAME THEORY*

that enabled him to escape from Hitler.[74] John Von Neumann would make a central contribution to economics more specifically by "develop[ing] a theory of [measurable] utility grounded in a set of assumptions about individual preferences."[75] The "Neumann–Morgenstern theory of utility ultimately received a share of attention from the economists" and one economist "in an early and laudatory review, actually went so far as to equate the maximization of 'expected profits or utilities' by individuals and firms with 'the principle of rational behavior' in economic theory."[76]

According to some scholars, the work of Von Neumann and Morgenstein was initially rejected by economists at Chicago.[77] Nonetheless, Von Neumann's work on utilities paved the way for "many of the keenest minds in America and throughout the world [to] devote[] to further delineating the science of rational choice"—which coincided with using "highly abstract, mathematically expressed conditions to which rational agency must conform and can be used to explain and predict the actions of agents supposed to be rational."[78]

Important evidence of Von Neumann's early influence at Chicago is the establishment of the Cowles Commission for Research in Economics. The Commission served as a hotbed for the spread of these ideas from 1939-1955. Commission economists applied his work with Morgenstein to, for example, the Federal Reserve's economic policies.[79] In the words of the Clifford Hildreth, who chronicled the work of the Commission, "publication of *Theory of Games* by Von Neumann and Morgenstern in 1944 was pivotal far several branches of economics" and "inspired lively interest and research," including "classical problems of monopoly, oligopoly, duopoly, and negotiations for unique commodities—

---

[74] *See generally* LEONARD, *supra* note 73, at 185- *See also* Till Düppe, *Siting the New Economic Science: The Cowles Commission's Activity Analysis Conference of June 1949*, 27 SCIENCE IN CONTEXT 453, 453–483 (2014).

[75] PAUL ERICKSON, THE WORLD THE GAME THEORISTS MADE 68 (2015); *see also* JOHN VON NEUMANN & OSCAR MORGENSTERN, THEORY OF GAMES AND ECONOMIC BEHAVIOR (1944)

[76] *Id.* at 69-70  Note that other economists resisted the influence of math, and voiced opposition to Von Neumann's influence  *Id.* at 70-71  *But see* Y  NARAHARI, GAME THEORY AND MECHANISM DESIGN 115 (2014) (discussing how utility theory became central to the notion of mechanism design, and devoting Chapter 8 to Von Neumann-Morganestein's utility theory)

[77] Düppe, *supra* note 74, at 474

[78] S  M  AMADAE, RATIONALIZING CAPITALIST DEMOCRACY: THE COLD WAR ORIGINS OF RATIONAL CHOICE LIBERALISM 7 (2003)

[79] Carl F  Christ, *History of the Cowles Commission 1932-1952*, in ECONOMIC THEORY AND MEASUREMENT: A TWENTY YEAR RESEARCH REPORT, COWLES FOUNDATION FOR RESEARCH IN ECONOMICS, YALE UNIVERSITY, https://cowles yale edu/research/publications/archive

*GAME THEORY*

houses, production facilities, etc."[80] One hang up was that Von Neumann and Morgenstein used probabilities, which at the time most economists viewed as inappropriate except to model "insurance, weather-based ventures, and gambling." But Abraham Wald at Columbia adapted game theory to apply to many economic problems, and "from 1944-54, Von Neumann and Morgenstern's expected utility representation. . . received substantial attention" and was widely accepted by thinkers building on it, including "many economists, statisticians, and other scientists."[81]

Game theory was also embraced by the RAND Corporation in Santa Monica, which lead to the invention of the Prisoner's Dilemma in 1950 by two scientists named Merrill Flood and Melvin Dresher.[82] The game was originally used as a military endeavor to model strategy, including the Cold War and the Korean War.[83] It was even used to model bombing exercises in the Pacific Theater of World War II, including at the Battle of Bismarck, where an ancestor of mine who was a top brass under General MacArthur used it to defeat a Japanese general who was sending a fleet through two possible alternative routes.[84]

Enter Anatol Rapoport, a mathematical biologist at Chicago and later Stanford. Rapoport felt that the results of the Prisoner's Dilemma should be tested experimentally and, in 1965, invented the strategy "Tit for Tat" to the Prisoner's Dilemma.[85] Rapoport "carried out extensive and empirical studies on the Prisoner's Dilemma" and described "an initial sequence of cooperation versus defection" where the strategy "tit-for tat . . . is one which apes the other player: one plays whatever the other played last time."[86] In 1979, Robert Axelrod, then a political scientist at the University of Michigan, decided to test the Tit for Tat hypothesis by inviting scholars to play at a computer tournament to see which program prevailed.[87]

---

[80] CLIFFORD HILDRETH, THE COWLES COMMISSION IN CHICAGO 1939-55, CENTER FOR ECONOMIC RESEARCH, DEPARTMENT OF ECONOMICS, UNIVERSITY OF MINNESOTA, DISCUSSION PAPER 225 (1985), COWLES FOUNDATION FOR RESEARCH IN ECONOMICS, YALE UNIVERSITY, https://cowles yale edu/sites/default/files/2022-08/misc-hildreth pdf
[81] *Id.* The Cowles Commission later decamped to Yale University where it remains today
[82] *Id.* at 19
[83] *Id.* at 193-226
[84] O G Haywood Jr , *Military Decision and Game Theory*, 4 JOURNAL OF THE OPERATIONS RESEARCH SOCIETY OF AMERICA 365, 365-71 (1954)
[85] Shirli Kopelman, *Tit for Tat and Beyond: The Legendary Work of Anatol Rapoport*, 13 NEGOTIATION AND CONFLICT MANAGEMENT RESEARCH 60 (2019)
[86] *Id.* at 67
[87] *Id.*

## GAME THEORY

The "Tit for Tat" initial answer to the Prisoner's Dilemma withstood the test of time for many years. It revolutionized game theory and had significant impacts on other fields as well, including evolutionary biology and economics. In Axelrod's seminal book *The Evolution of Cooperation*, a chapter co-authored by William D. Hamilton argued that the "Tit for Tat" model could apply to "interactions between two bacteria or interactions between two primates," and that it could also apply to "the interactions between a colony of bacteria and, say, a primate serving as a host" using the theoretical models developed by Axelrod earlier in the book.[88] They observed that not only could the model apply to animals and to microbes, but that it could also be used to study human interactions and problems ranging from the "causes of cancer" to "certain kinds of genetic defects," including, for example, Down's Syndrome.[89] Study of what is known as the Iterated Prisoners Dilemma (IPD)—in which the game is played repeatedly—was also essential to the study of cooperation.[90] The IPD and other 2 x 2 repeated games later became incorporated into the work of more recent thinkers in the Chicago School as a form of classical law and economics, and behavioral law and economics.[91]

The Tit for Tat view, while not the only accepted answer to Prisoner's Dilemma from the perspective of the Chicago School,[92] is still reflected in many Chicago School models of the legal system. As Professor Picker explains, "[t]he power of the Prisoner's Dilemma comes from the incongruence between private benefit and the collective good. Individually rational decisionmaking leads to collective disaster. The Prisoner's Dilemma is thus often seen as one of the main theoretical justifications for government intrusion into private decisionmaking."[93] Or, in Judge

---

[88] AXELROD, *supra* note 22, at 126

[89] *Id.* at 137-38

[90] *Id.*

[91] Picker, *supra* note 4, at 5; *see, e.g.,* CASS R SUNSTEIN, AFTER THE RIGHTS REVOLUTION: RECONCEIVING THE REGULATORY STATE 49-51 (Harvard Univ Press, 1990).

[92] *See* Picker, *supra* note 4, at 20 ("I found even more surprising the notion that having a Prisoner's Dilemma handy might actually help solve collective action problems, rather than create them, and that this should make us cautious in relying on the Prisoner's Dilemma to justify legal intervention "); BAIRD, GERTNER, & PICKER, *supra* note 8, at 170-173 (discussing "slightly different version of tit-for-tat if the strategy is to be subgame perfect" and discussing other solutions such as "Pareto-optimal equilibria [that] rely on very severe punishment" and concluding "[n]one of the existing methods of narrowing the possible equilibria proves completely satisfactory Nevertheless, the models do show that repetition itself creates the possibility of cooperative behavior The mechanism that supports cooperation is the threat of future noncooperation for deviations from cooperation ")

[93] Picker, *supra* note 4, at 5 However, Picker clarifies that "we need to know much about the quality of government decisionmaking before we can summarily abandon private decisionmmaking " Picker,

## *GAME THEORY*

Posner's view, the legal system substitutes to deter bad behavior such that "[t]he severity of most [] penalties is an indication that society considers [the] activity [be it speeding or murder] to be socially very costly" and prevention of future wrongdoing requires penalties, even though these penalties may discourage some lawful conduct."[94] As Judge Posner notes with his co-author William Landes, this view is true not only of the criminal justice system, but also of the civil system, including lawsuits to prevent large-scale wrongdoing by corporations.[95]

### A. Tort Law

Scholars in the Chicago School, for example, use a "due care" game that is a variation on the Prisoner's Dilemma as a foundational model for the tort law system of the United States.[96] In a lecture explaining the basics of game theory, Professor Picker proposes the following payoff matrix in a 2 x 2 game modeling how pedestrians and motorists should behave in a "legal regime of no liability."[97]

---

*supra* note 4, at 5   He also notes the Prisoner's Dilemma needs to be paired with other games and that "it is a mistake to suggest that a Prisoner's Dilemma may arise in a particular context and to use that to justify legal invention  The larger game structure must be understood, as these stylized games suggest " *Id.* Picker does, however, sense the flaws with his model of the Prisoners' Dilemma, concluding the piece that that pairing the Prisoners' Dilemma with other games "counterintuitive[ly]" may "actually *help* the player to achieve the best outcome " *Id.*
[94] Ehrlich & Posner, *supra* note 7, at 272
[95] *See* William M  Landes & Richard A  Posner, *Symposium: The Positive Economic Theory of Tort Law*, 15 GEORGIA L  REV  851, 857-58 (1982)
[96] BAIRD, GERTNER, & PICKER, *supra* note 6, at 47-49 ("[R]ecent work on torts have started to return to the explicit use of game theory  Examples include Orr (1991) and Chung (1992)  Both argue for a comparative negligence standard on the basis of dominant strategies for both parties  The sharing rule for comparative negligence cases that we introduce in the text—and variations of it—have been mentioned or advocated in a variety of articles ")
[97] Picker, *supra* note 4, at 7

*GAME THEORY*

|  | Motorist | |
|---|---|---|
|  | No Care | Due Care |
| **Pedestrian** No Care | *-100, 0* | -100, -10 |
| **Pedestrian** Due Care | -110, 0 | -20, -10 |

**Payoffs: (Pedestrian, Motorist)**

*Figure 2: Payoff Matrix in a "due care" game where a motorist and a pedestrian where "[i]f an accident takes place between the motorist and the pedestrian, the motorist and her car will not be hurt, but the pedestrian will of course suffer harm. Assume that we can represent the harm to the pedestrian as a dollar amount and set that amount at $100." Both the Pedestrian and the Motorist have the option to play "No Care" or "Due Care" and rewards are again listed with the row player's (Pedestrian's) reward first. Note that the Motorist is better off playing "No Care" irrespective of Pedestrian's move, while the Pedestrian's strategy depends on the Motorist's actions. Figure reproduced from Randal C. Picker,* An Introduction to Game Theory and the Law *at 2 (University of Chicago Coase-Sandor Institute for Law & Economics Working Paper No. 22, 1994).*

  Professor Picker, both in his article and in his canonical book *Game Theory and the Law*, co-authored by Baird, and Gertner, claim the game demonstrates that a "strictly dominant strategy" for the motorist is never to take due care in a legal regime of no liability.[98] Indeed, "taking care costs the motorist $10 and provides no benefit to the motorist in return. The motorist always does better by not taking care than by taking care."[99] The incentives of the pedestrian are a little harder to predict. Unlike in the Prisoner's Dilemma, "the pedestrian lacks a dominant strategy because either course of action could be better or worse than the other depending upon what the motorist does."[100] But given the motorist's strategy, the pedestrian "is better off not exercising care either."[101] The point? "[T]he language of game theory states a familiar insight from law and economics, the world in which without tort law, parties tend to take less due care because they do not fully internalize the costs of their actions."[102] The game theoretic modeling of this scenario, is, according to

---

[98] Picker, *supra* note 4, at 8 *See also* BAIRD, GERTNER, PICKER, *supra* note 6, at 13-14
[99] Picker, *supra* note 4, at 8 *See also* BAIRD, GERTNER, PICKER, *supra* note 6, at 13-14
[100] Picker, *supra* note 4, at 8
[101] *Id.*
[102] *Id.* (citing WILLIAM LANDES & RICHARD POSNER, THE ECONOMIC STRUCTURE OF TORT LAW (1987)) *But see* Jack Balkin, *Book Review: Too Good to Be True? The Positive Economic Theory of Law*, 87 COLUM L REV 1447, 1447-48 (1987) (noting "a major failing of the book is its reductionism, a

*GAME THEORY*

Picker, what underlies the need for the tort system and its concept of "contributory negligence."[103]

This model underlies not only the notion of "due care" for traffic accidents and the like, but also larger concerns in products liability cases that the tort system is founded upon. As Professor Landes and Judge Posner explain, "in a world of no liability, [manufacturing] output will exceed the optimal output. . . and the manufacturer will fail to take due care."[104] Although according to Landes and Posner, the marketplace will correct many failures on the part of the manufacturer—especially when consumers are adequately warned of possible product defects—in some instances the power of the market is not enough, as in the case of "*highly dangerous products*."[105]

In an article, Judge Posner uses the example of Ford Motor Company "knowingly install[ing] an abnormally and dangerously fragile gas tank in its Pinto automobiles without warning customers or offering a compensation price reduction."[106] In that case, Posner explains, "negative altruism" would cause consumers to "refuse to buy Ford products in the future."[107] But, in Landes' and Posner's view, the "empirical evidence tends to support the proposition that tort law deters too."[108] The economic view of tort law, in their eyes, provides empirical support for the Aristotelian view of the tort system as a form of "corrective justice," namely "to restore to a person that which has wrongfully been taken from him, rather than to improve allocation of resources."[109] And Judge Posner also supervised a book series called *Economic Approaches to Law* that includes a book called *Game Theory and the Law* using utility functions to explain various aspects of the legal system, including liability rules for contributory negligence.[110]

But examples from the real world indicate that the tort system and the corrective justice model it tracks have widely failed. Take, for

---

reductionism that occurs on two levels: its attempts to view tort law as animated by a single purpose—efficiency, and its attempts to envision all human behavior as market behavior ")
[103] Picker, *supra* note 4, at 11
[104] *See* Landes & Posner, *supra* note 102, at 293; *but see id.* at 276-277 (noting that at first blush no liability would result in an optimum distribution and could be "left to the marketplace to set" but noting how this conclusion "may not be right, because transaction costs may actually be high")
[105] *Id.* at 291-99
[106] Posner, *Retribution and Related Concepts*, *supra* note 7, at 80
[107] *Id.* at 78
[108] Landes & Posner, *supra* note 95, at 858
[109] Landes & Posner, *supra* note 95, at 857-58; *see also* Helen E White, *Making Black Lives Matter: Properly Valuing the Rights of the Marginalized in Constitutional Torts*, 128 YALE L J 1742, 1767, 1775 (2019)
[110] *See* GAME THEORY AND THE LAW 395-399 (Ed Eric B Rasmussen, 2008)

*GAME THEORY*

example, the opioids epidemic. A FOIA matter I served as co-lead counsel on produced documents showing that the FDA knew painkillers were addictive for decades without taking sufficient action.[111] Plaintiffs' attorneys suing the pharmaceutical companies and distributors allegedly responsible for the opioids epidemic argued that due to "the proliferation of opioid pharmaceuticals . . . life expectancy for Americans decreased for the first time in recorded history" and—"[d]rug overdoses" as of 2019 were "the leading cause of death for Americans under 50."[112] Strikingly, these deaths occurred despite the fact that the company recognized as initiating the opioids epidemic[113]—Purdue Pharma—was investigated, indicted, and pled guilty in federal district court in Virginia in 2007 to

---

[111] *See* Abby Goodnough & Margot Sanger-Katz, *As Tens of Thousands Died, F.D.A. Failed to Police Opioids*, N Y TIMES (Dec 31, 2019), https://www nytimes com/2019/12/30/health/FDA-opioids html (describing documents produced over the course of my negotiations with the FDA representing researchers who had been stonewalled for years in their request) *See generally* BEN GOLDACRE, BAD PHARMA: HOW DRUG COMPANIES MISLEAD DOCTORS AND HARM PATIENTS (2012) (exploring problems with the pharmaceutical industry and advocating for more transparency as the solution) Many scholars advocate for the use of transparency to combat problems such as the opioids epidemic, and that with proper labeling or public awareness, problems like these will be ameliorated *See* Amy Kapczynski, *The Law of Informational Capitalism*, 129 YALE L J 1460, 1508 (2020) (reviewing SHOSHANA ZUBOFF, THE AGE OF SURVEILLANCE CAPITALISM: THE FIGHT FOR A HUMAN FUTURE AT THE NEW FRONTIER OF POWER (2019) & JULIE E COHEN, BETWEEN TRUTH AND POWER: THE LEGAL CONSTRUCTIONS OF INFORMATIONAL CAPITALISM (2019)) A recent paper by Amy Kapczynski and her colleague Christopher Morton at Columbia argues that we need more than transparency, we need, in the context of drugs, *data publicity* and we need the FDA working with independent scientists to catch safety and efficacy issues that might be missed by the agency *See* Amy Kapczynski & Christopher Morten, *The Big Data Regulator: Why and How the FDA Can and Should Disclose Confidential on Prescription Drugs*, 109 CALIF L REV 493 (2021) But the "transparency trap" as I refer to it cannot explain why widespread social ills, such as child abuse and the Catholic Church, or the Black Lives Matter protests and police killings, have not ended these widespread social problems *See* TROUBLING TRANSPARENCY: THE HISTORY AND FUTURE OF FREEDOM OF INFORMATION (David E Pozen & Michael Schudson eds , 2018); Margaret B Kwoka, *FOIA Inc* , 65 DUKE L J 1361 (2016); MATT CARROLL ET AL , BETRAYAL: THE CRISIS IN THE CATHOLIC CHURCH (2002)

[112] Complaint ¶ 6, School Board of Miami Dade v Endo, No 1:19-cv-24035 (S D Fla Sept 30, 2019) According to the docket, the case was transferred to the multidistrict litigation occurring in Ohio overseen by Judge Dan Poster that consolidates thousands of individually filed lawsuits related to the opioids epidemic and that is still pending *See id.*, ECF No 8

[113] Opioids have been used—and abused—for millennium in early Egypt as well as China *See* Nick Miroff, *From Teddy Roosevelt to Trump: How drug companies triggered an opioid crisis a century ago*, Wash Post, Oct 17, 2017, https://www washingtonpost com/news/retropolis/wp/2017/09/29/the-greatest-drug-fiends-in-the-world-an-american-opioid-crisis-in-1908/ The DEA has regulated prescription opioids since 1970 as "Schedule II" controlled substances Complaint ¶ 89, *School Board of Miami Dade* Although certain prescription drugs that were used—and abused—were developed such as "Percodan, Percocet, and Vicodin," these had "relatively low opioid content" and it was not until the 1996 launch of Oxycontin that the "modern opioid[s] epidemic" began *Id.* ¶ 90 Notably, "[t]he weakest OxyContin delivers as much narcotic as the strongest Percocet, and some OxyContin tablets delivered sixteen times that " *Id.* ¶ 91 And, "[p]rior to Purdue's launch of OxyContin, no drug company had ever promoted such a pure, high-strength Schedule II narcotic to so wide an audience of general practitioners " *Id.* ¶ 112

## GAME THEORY

felony misbranding of its drug OxyContin "with the intent to defraud and mislead the public."[114] According to a 2007 statement from the Department of Justice accompanying the plea agreement—which involved admissions of guilt from Purdue Pharma—Purdue pitched OxyContin as a "wonder" drug that would bring "hope and relief to millions" of patients suffering from chronic pain and marketed as a drug bringing long-acting relief that was less addictive than existing painkillers.[115] But, as Purdue was well aware, OxyContin was "nothing more than pure oxycodone—a habit forming narcotic derived from the opium poppy."[116]

Purdue's "aggressive marketing campaign," falsely led health care providers to believe that OxyContin was more difficult to abuse by crushing it for intravenous use, falsely stated the drug was less addictive than immediate-release opioids, falsely created the impression the drug did not lead to "peaks and troughs" and did not produce state of euphoria, falsely told health care providers there were no withdrawal effects and that patients would not develop tolerance, and falsely suggested that OxyContin could be used to "weed out" addicts due to the above.[117] The marketing campaign led OxyContin to "be[come] the new pain medication of choice for many doctors and patients" with "skytocket[ing]" sales that made "billions for Purdue and millions for its top executives" even while leading to a wave of addiction, drug-related deaths, and crime directly-attributable to the drug.[118] As a result of the plea agreement in 2007, Purdue agreed to "pay over $600 million in criminal and civil penalties" and to undergo "independent monitoring and an extensive remedial action program," and three Purdue executives pled guilty to misdemeanors and "agreed to pay $34.5 million in penalties."[119]

Yet, according to plaintiffs' attorneys in now-pending multidistrict litigation in Ohio, the 2007 criminal and civil penalties levied against Purdue and three top executives appeared to have had no deterrent effect on Purdue itself or on other pharmaceutical companies. Tellingly, Purdue's sales for opioids in 2015 totaled $3 billion—"represent[ing] a fourfold increase from its 2006 sales of $800 million."[120] Building on the

---

[114] STATEMENT OF UNITED STATES ATTORNEY JOHN BROWNLEE ON THE GUILTY PLEA OF THE PURDUE FREDERICK COMPANY AND ITS EXECUTIVE FOR ILLEGALLY MISBRANDING OXYCONTIN, U.S. DEP'T OF JUSTICE, WESTERN DISTRICT OF VIRGINIA 3 (May 10, 2007)
[115] *Id.*
[116] *Id.*
[117] *Id.*
[118] *Id.*
[119] *Id.*
[120] Complaint ¶ 114, *School Board of Miami Dade*

*GAME THEORY*

market for chronic pain—a market many doctors were deeply sympathetic to—competitors copied Purdue's tactics and used the active ingredient in OxyContin to manufacture competing drugs that eventually displaced OxyContin as market leaders.[121] Indeed, according to data obtained by the *Washington Post,* "the number of pills made with oxycodone—the main ingredient in OxyContin—rose from 2.5 billion in 2006 to 4.5 billion in 2012, an 80 percent increase."[122]

Despite Purdue's original guilty plea and its most recent guilty plea in November 2020,[123] systemic efforts to halt the opioids epidemic have remained elusive. This is despite the persistence of the district court tasked with handling the multidistrict litigation brought by state attorneys general as well as by affected cities, counties, and Indigenous tribes across the United States. Judge Dan Polster of the Northern District of Ohio has been tasked with overseeing multidistrict litigation—which as September 2019 "totaled more than 2,000 individual actions."[124] Judge Polster pushed aggressively for settlement, appearing to view himself as personally responsible for ending the opioids epidemic, while noting during the first hearing that "about 150 Americans are going to die today, just today, while we're meeting."[125] Judge Poster's methods have been unconventional—to say the least. In a *New York Times* profile, Judge Poster made clear that he did not intend to use discovery or "preside over years of 'unraveling complicated conspiracy theories'" when he viewed as undercutting efforts to solve a massive social problem when the parties had "at least 80 percent of the information they need to negotiate" and "the longer litigation continues, . . . the more entrenched each side can become."[126]

Yet Judge Polster's aggressive stance towards encouraging settlement (an approach that had led him to successfully resolve other cases)

---

[121] *Id.* ¶ at 118-25; Sari Horwitz, Scott Higham, Dalton Bennett & Meryl Kornfield, *'SELL BABY SELL!': Unsealed documents in opioids lawsuit reveal inner workings of industry's marketing machine*, WASH POST, Dec 6, 2019, https://www washingtonpost com/graphics/2019/investigations/opioid-marketing/
[122] Horwitz, Rich & Higham, *supra* note 121
[123] DEP'T OF JUSTICE, OPIOID MANUFACTURER PURDUE PHARMA PLEADS GUILTY TO FRAUD AND KICKBACK CONSPIRACIES 64 (Nov 24, 2020), https://www justice gov/opa/pr/opioid-manufacturer-purdue-pharma-pleads-guilty-fraud-and-kickback-conspiracies
[124] Memorandum Opinion Certifying Negotiating Class at 1, In re: National Prescription Opiate Litigation, No 1:17-MD-2804 (N D Ohio Sept 11, 2019), ECF No 2590
[125] Jan Hoffman, *Can This Judge Solve the Opioid Crisis?*, N Y TIMES, (Mar 5, 2018), https://www nytimes com/2018/03/05/health/opioid-crisis-judge-lawsuits html
[126] *Id.*

*GAME THEORY*

did not yield settlement in this matter as of the time of this writing.[127] Rather, the parties objected to his unorthodox methods—including challenging his use of the Federal Rules of Civil Procedure—leading to him being rebuked by the Sixth Circuit for failing to adhere to the law as written.[128] And while Judge Polster's tactics did lead Purdue to decide not to market OxyContin to prescribers,[129] and also led to findings that CVS and Walmart were liable for their role in the epidemic,[130] Purdue filed for bankruptcy in the Southern District of New York in an attempt to shield itself from liability in Ohio.[131]

    The New Hampshire State Attorney General in March 2022 announced a sweeping settlement with Purdue that would lead to a payment from the Sackler family of "up to $6 billion if certain conditions are met."[132] But several other state attorneys general objected and the district court judge "rejected the deal over concerns of expansive protections shielding the wealthy Sackler family from future lawsuits."[133] As a result, the settlement is the subject of an ongoing appeal to the Second Circuit, which held oral argument and appeared "hesitant. . . to revive a settlement plan . . . that would shield individual members of the Sackler family from current and future civil lawsuits over their role in promoting the opioid epidemic."[134] In sum, the opioids epidemic was preventable tragedy caused by humans, and the tort system proved wholly unable to correct it—even when enough of the information was available to approximately allocate liabilities—an outcome that defies the game theoretic modeling of the Chicago School.

---

[127] *See* Order, Schedule for Track Three Abatement Phase, In re: National Prescription Opiate Litigation, No 1:17-MD-2804 (Dec 10, 2020)
[128] In Re: National Prescription Opiate Litigation, No 20–3075 at *2 (6th Cir April 15, 2020)
[129] Hoffman, *supra* note 125
[130] *See* Abatement Order, In re: National Prescription Opiate Litigation, No 1:17-MD-2804 (Aug 17, 2022)
[131] *Id.*
[132] NEW HAMPSHIRE DEP'T OF JUSTICE, ATTORNEY GENERAL FORMELLA ANNOUNCES UP TO $6 BILLION NATIONAL SETTLEMENT WITH PURDUE PHARMA AND SACKLERS; NEW HAMPSHIRE TO RECEIVE $46 MILLION IF AGREEMENT APPROVED (Mar 3, 2022), https://www doj nh gov/news/2022/20220303-settlement-purdue-pharma-sacklers htm
[133] *See* Josh Russell, *Second Circuit looks askance at $6 billion OxyContin settlement*, Courthouse News Service (Apr 29, 2022), https://www courthousenews com/second-circuit-looks-askance-at-6-billion-oxycontin-settlement/
[134] *Id.*

*GAME THEORY*

B. Criminal Law

The criminal law system is also rationalized by the Chicago School and classical law and economics. In a canonical article, Judge Posner explained that "the major function of criminal law in a capitalist society is to prevent people from bypassing the system of voluntary, compensated exchange—the 'market,' explicit or implicit—in situations where, because transaction costs are low, the market is a more efficient method of allocating resources than forced exchanged. Marketing bypassing in such situations is inefficient—in the sense in which economists equate efficiency with wealth maximization—no matter how much utility it may confer on the offender."[135] Judge Posner acknowledges that "deriv[ing] basic criminal prohibitions from the concept of efficiency" is a "controversial endeavor" but that in his view, "the role of the criminal law in discouraging market bypassing is obscured by the fact that the market transaction that the criminal bypasses is usually not a transition with [their] victim."[136]

Even more controversially, Posner suggests that "[g]iven the economist's definition of 'value,' even if the rapist cannot find a consensual substitute (and one such substitute, prostitution, is itself illegal), it does not follow that he values the rape more than the victim disvalues it. There is a difference between a coerced transaction that has no consensual substitute and one necessary to overcome the costs of consensual transactions; only the second can create wealth, and therefore be efficient. Indeed, what the argument boils down to is that some rape is motivated in whole or in part by the negative interdependence of the parties' utilities, and this, as I have argued in connection with crimes of passion, is no reason for considering the act efficient."[137]

---

[135] Posner, *An Economic Theory of the Criminal Law*, *supra* note 7, at 1194-95
[136] *Id.*
[137] Posner makes even more choice statements about rape, continuing:

> As with my earlier discussion of crimes of passion, it is important not to take too narrow a view of market alternatives  Supposing it to be true that some rapists would not get as much pleasure from consensual sex, it does not follow that there are no other avenues of satisfaction open to them  It may be that instead of furtively stalking women they can obtain satisfactions from productive activities, that is, activities in which other people are compensated and thus derive benefits  This is an additional reason to think that the total wealth of society would be increased if rape could be completely repressed at a reasonable cost  All this may seem to be a hopelessly labored elucidation of the obvious, that rape

## *GAME THEORY*

The use of criminal law to extract vengeance is also something that makes economic sense, according to the Chicago School. Judge Posner wrote a 1997 claiming that the influence of "evolutionary biology" "enable[d] the concept of rationality to be enlarged to cover phenomena (not only fairness but at least one of the cognitive quirks, the sunk-costs fallacy) that [could be considered] as irrational."[138] Judge Posner continued that evolutionary biology's treatment of altruism and revenge "are dimensions of rationality that I have been writing about for many years."[139] In Posner's view, for example, evolutionary biology supports long prison sentences as an effective mechanism of deterrence because "lengthening a prison sentence from twenty to twenty-five years will increase its disutility (in 'present value' terms, that is, as reckoned by the defendant when [they] [are] deciding whether to commit a criminal act that would expose [them] to punishment) by much less than twenty-five percent; at a discount rate . . . the increase will only be about six percent."[140]

Judge Posner also explained that, in evolutionary biology terms, revenge is rational and should be supported by the legal system, which stands in the shoes of the wronged individual or group. In his words, the appetite for vengeance "is between ex post and ex ante rationality" because "[h]aving an unshakable commitment to retaliate may be ex ante rational by lowering the risk of being a victim of aggression, even though, if the risk materializes, acting on the commitment will then (that is, ex post) become irrational."[141] Judge Posner argues revenge "[is] needed for deterrence and hence survival in the state of human society before there were any formal legal or political institutions, and thus before it was possible to make a legally enforceable commitment to retaliate against an aggressor, was an instinctual commitment to retaliate" and while

---

       is a bad thing; but I think it useful to point out that economic analysis need not
       break down in the face of such apparently noneconomic phenomena as rape

*Id.* Most feminists and scholars of women's rights would take decided issue with Judge Posner's characterization of rape as explained by economics *See* Martha F Davis & Susan J Kraham, *Protecting Women's Welfare in the Face of Violence*, 22 FORDHAM URB L J 1141, 1144 (2011) ("Each year, between three and four million women in this country are battered by husbands, partners[,] and boyfriends Half of these women are beaten severely and, in 30% of the domestic violence incidents reported, the assailants used weapons "); Lisa R Pruitt, *Place Matters: Domestic Violence and Rural Difference*, 23 WIS J L GENDER & SOC'Y 347 (2008) (noting counterintuitive findings on urban and rural domestic violence to show that power differences in education and status play a huge role in domestic abuse)
[138] Posner, *Rational Choice, Behavioral Economics, and the Law, supra* note 7, 1563
[139] *Id.* at 1568
[140] *Id.* at 1563
[141] *Id.* at 1563

## *GAME THEORY*

"sometime retaliation end[s] in disaster" the "people who were endowed with an instinct to retaliate would have tended to be more successful in the struggle for survival than others."[142]

Take the view of punishment espoused by Judge Posner as rooted in "the *lex talionis* of early Roman law, the 'eye for eye precept in the Old Testament" that "counts among its distinguished philosophical exponents Immanuel Kant" and John Rawls, who have argued that it is "morally fitting that a person who does wrong should suffer in proportion to his guilt."[143] According to Posner, harsh punishments, including prison and the death penalty are warranted. Imprisonment "both reduces the criminal's future wealth, by impairing his lawful job prospects, and imposes disutility on people who cannot be made miserable enough by having their liquid wealth, or even their future wealth confiscated."[144] And, in crimes like murder, where the cost to the "victim approach infinity, even life imprisonment may not impose costs on the murderer equal to the victim."[145]

Posner notes that "there is no realistic method of preserving marginal deterrence for every crime, though medieval law tried" by using gruesome techniques such as "boiling in oil" to punish "murder by poisoning" since "poisoners were especially difficult to apprehend in those times, a heavier punishment than that prescribed for ordinary murderers was (economically) indicated."[146] Judge Posner goes on to discuss flogging by parents, noting his main objection is not that "inflicting physical pain" is "disgusting" but rather "[j]ust to inflict a momentary excruciating pain with no aftereffects might be a trivial deterrent," while to inflict a level of pain that "would be the equivalent to five years in prison would require measures so drastic that they might endanger the life, or destroy the physical or mental health, of the offender."[147] Posner concludes that there are other ways to punish "other than by varying the length of imprisonment" through "[s]ize of the prison cell, temperature, and quality of food" and noting "[m]inimum security prisons are more comfortable than intermediate security prisons, and the latter are more comfortable than maximum security prisons."[148]

---

[142] *Id.*
[143] Posner, *supra* note 55, at 71
[144] Posner, *An Economic Theory of the Criminal Law*, *supra* note 7, at 1209
[145] *Id.* at 1210
[146] *Id.*
[147] *Id.*
[148] *Id.* at 1211

## *GAME THEORY*

So too others in the Chicago School such as Judge Easterbrook have rationalized the criminal justice system using a utility-based approach with supply and demand curves implicitly resting on fundamental assumptions of game theory. Judge Easterbrook notes that that the systematic abuses of the system are merely its costs, as the Appendix to his article demonstrates in terms of its approach to settlement and plea bargaining.[149] In Judge Easterbrook's words: "Pervasive discretion and pervasive injustice are different things. The discretion in criminal procedure does not produce chaos and systematic arbitrariness. It yields, instead, the order of the marketplace, coordination of the acts of many thousands of people through a price system. The features of criminal procedure I have examined in this paper are best understood as if designed to facilitate a market assessment and imposition of the price of crime. At every turn, the legal doctrines track would be desirable in a system constructed with minimum allocative inefficiency. I do not say that all doctrines were that they do have this effect. . . . Given their existence, the legal doctrines seem reasonably well designed to squeeze the maximum deterrence out of funds available to courts and prosecutors. This is largely a positive conclusion though I am content with its normative implications."[150]

But as the abolitionist movement has noted, punishment empirically fails to help most victims. This is due to system failures that show the system is racialized[151] and fails to protect victims who are persons of color. And even when punishment "works" it does not relieve the intense trauma victims experience. Roxanna Altholz at the UC Berkeley International Human Rights Clinic notes that in partnering with the victims of unsolved crimes in Oakland, California, there are significant problems with the police and their relationship to the communities they are dedicated to serving in terms of bringing closure to victims of violent crimes.[152] Although Oakland has one of the highest crimes rates relative

---

[149] Frank H Easterbrook, *Criminal Procedure as a Market System*, 12 J OF LEGAL STUDIES 289, 330-32 (1983)
[150] *Id.*
[151] A complete treatment of racial disparities in criminal justice is beyond the scope of this article *See generally* MICHELLE ALEXANDER, THE NEW JIM CROW: MASS INCARCERATION IN THE AGE OF COLORBLINDNESS (10th ed 2020)
[152] ROXANNA ALTHOLZ, INT'L HUM RTS L CLINIC, U C , BERKELEY SCH L , LIVING WITH IMPUNITY: UNSOLVED MURDERS IN OAKLAND AND THE HUMAN RIGHTS IMPACT ON VICTIMS' FAMILY MEMBERS, 10, 23 (2020), https://www law berkeley edu/wp-content/uploads/2020/01/Living-with-Impunity pdf; Anthony A Braga et al , *Oakland Ceasefire Impact Evaluation: Key Findings*, CITY OAKLAND 8 (Aug 10 2018), https://cao-94612 s3 amazonaws com/documents/Oakland-Ceasefire-Evaluation-Final-Report-May-2019 pdf (emphasizing finding that Oakland police must treat residents with "respect" and "dignity")

*GAME THEORY*

to other cities in California, it is characterized by glaring racial disparities—for example "police made arrests in approximately 40% of the Oakland homicides involving black victims and approximately 80% of the homicides involving white victims" and the "Oakland Police Department (OPD) has over 2,000 cold homicide cases on its books."[153] To the extent that it is a city with a high crime rate, Oakland's "backlog" of "unsolved murder cases," is, in fact "comparable" to the norm.[154]

Indeed, statistics on prison leading to further violence are striking. Leading scholars such as Jeffrey Fagan and Tracey Meares have demonstrated that "it is possible that punishment may have backfired" because "there is growing evidence from several small-scale studies that incarceration may have iatrogenic effects"—namely it has caused further crime instead of preventing it.[155] A systematic review to assess the effect of prison on recidivism found "[a]cross all comparisons . . . incarceration resulted in a 7% increase in recidivism compared with a community sanction."[156] Subsequent research that "focused on high-quality research designs" should the "effect associated with imprisonment jumped to 11%" and that "even when the weighted mean effect size was calculated the iatrogenic effect of imprisonment remained, with custodial sanctions associated with an 8% increase in recidivism."[157] A final systematic review "meta-analyzed 57 studies" and discovered that prison is "[i]nconsistent with deterrence theory, [and] harsher conditions were associated with recidivism."[158] Thus, the Chicago School's emphasis on punishment and harsh prison conditions is not borne out by the empirical literature on punishment as younger voices within the Chicago School have noted.

## II. WIN-WIN SOLUTIONS

Part I showed that the Chicago School is not structurally equipped to promote "win win" solutions. This Part shows that all hope is not lost. While a major redesign is not on the minds of many, this

---

[153] Altholz, *supra* note 152, at 1
[154] *Id*. at 6
[155] Jeffrey Fagan & Tracey L Meares, *Punishment, Deterrence and Social Control: The Paradox of Punishment in Minority Communities*, 6 OHIO ST J CRIM L 173, 176 & n 11 (2008) (collecting studies): *see also* Francis T Cullen, Cheryl Lero Jonson, and Daniel S Nagin, *Prisons Do Not Reduce Recidivism: The High Cost of Ignoring Science*, 91 THE PRISON J 48S, 54S-58S (2011)
[156] Cullen et al , *supra* note 155, at 54S-58S
[157] *Id.*
[158] *Id.*

## *GAME THEORY*

Article begins the call to imagine a blueprint for a new legal movement built on pure cooperation, not defection. This is no small feat—but to start imagining we must think small. This Article borrows the famous metaphor of the cathedral—which connotes a vision of law as a sacred space—from the canonical article by Guido Calabresi and A. Douglas Melamed on the difference between property rules and liability rules.[159] The gap between recognizing these preliminary insights and translating them into our legal school effectively—the way an architect might translate a blueprint of a cathedral—is, to say the least, enormous. Cathedrals sometimes take hundreds of years to build and they require not only an architect, but also teams of engineers, builders, masons, and others to help turn a two-dimensional image into a reality. This Article calls on legal scholars and activists to take the incredibly rough sketch and flesh it out over the coming years. Once we understand these basic design features, we must next understand how to maintain and scale cooperation at a national and eventually international level.

### A. New Answers to Old Dilemmas in Math and Nature

Recent innovations in physics, engineering, and evolutionary biology give new hope for rejecting the outdated thinking that the best solution for cooperation is premised on the philosophy of an "Eye for an Eye" or "Tit for Tat." Indeed, these innovations show the potential of win-win solutions to lead to mutual benefit and maximal cooperation. "Tit for Tat" emerges when the Prisoner's Dilemma is played repeatedly—where the players encounter each other more than once. These longer-term interactions change the theoretical outlook of the game, as well as the practical insights that can be derived from it to help promote cooperation on the ground. But the strategy of "Tit for Tat," and the conclusions reached from it, was flawed when it won the tournament in which it played, and it is equally flawed now insofar as it does not represent the correct theoretical answer to the question of how to maximize points within the Iterated Prisoner's Dilemma.[160]

Early critics intuited that "Tit for Tat" was an imperfect solution that was highly variable, pointing out that "the quality of the [Tit for Tat]

---

[159] *See* Guido Calabresi & A. Douglas Melamed, *Property Rules, Liability Rules, and Inalienability: One View of the Cathedral*, 85 HARV. L. REV. 1089, 1090 n.2.
[160] Even Axelrod himself intuited that "Tit for Tat" was flawed, stating: "A major lesson of this tournament is the importance of minimizing echo effects in an environment of mutual power. When a single defection can set off a long string of recriminations and counterrecriminations, both sides suffer

*GAME THEORY*

strategy depends entirely on the quality of the environment" and thus should not always hold true.[161] As the work of subsequent pioneers showed conclusively in 2012, there is another class of strategies that not only defeats "Tit for Tat," but also defeats other improved models built on "Tit for Tat" and correctly answers the age-old paradox. These strategies allow a player with a "theory of the mind" to defeat players without one.[162] These pioneers, computational biologist Professor William Press, working from the University of Texas, Austin, and physicist-mathematician Professor Freeman Dyson of Princeton, "shocked the world of game theory"[163] by building on the lost insight of one of the other programs from the original tournament called DOWNING.[164]

Press and Dyson showed an entirely new class of strategies they labeled Zero Determinant (ZD) strategies.[165] Like in the original Axelrod tournament, ZD strategies are not strangers—they will encounter each other (or variants of the other) in an iterated game.[166] Like "Tit for Tat," ZD strategies only use information from the outcome of the most recent previous interaction with the other player (namely, they have a "memory" of 1).[167] However, rather than a deterministic framework of "if previous outcome was X, always do Y," ZD strategies are "stochastic"—they assign probabilities to their next move based upon the

---

A sophisticated analysis of choice must go at least three levels deep to take account of these echo effects The first level of analysis is the direct effect of a choice This is easy, since a defection always earns more than a cooperation The second level considers the indirect effects, taking into account that the other side may or may not punish a defection This much of the analysis was certainly appreciated by many of the entrants But the third level goes deeper and takes into account the fact that in responding to the defections of the other side, one may be repeating or even amplifying one's own previous exploitative choice Thus[,] a single defection may be successful when analyzed for its direct effects, and perhaps even when its secondary effects are taken into account But the real costs may be in the tertiary effects when one's own isolated defections turn into unending mutual recriminations Without their realizing it, many of these rules wound up punishing themselves " AXELROD, *supra* note 22, at 61

[161] Peter Huber, *Competition, Conglomerates, and the Evolution of Cooperation*, 93 YALE L J 1147, 1158 (1984) (reviewing AXELROD, *supra* note 26)

[162] William H Press & Freeman J Dyson, *Iterated Prisoner's Dilemma Contains Strategies That Dominate Any Evolutionary Opponent*, 109 PROC NAT'L ACAD SCI U S AM (PNAS) 1049, 1049 (2012) *See also* Stewart & Plotkin, *supra* note 17, at 10135

[163] Baillie, *supra* note 68

[164] AXELROD, *supra* note 22, at 56-57; Douglas R Hofstadter, *Metamagical Themas: Computer Tournaments of the Prisoner's Dilemma Suggest How Cooperation Evolves*, 248 SCI AM 16, 22 (1983) (noting that by Alexrod's own calculations a program called REVISED DOWNING should have won)

[165] Press & Dyson, *supra* note 162, at 10409

[166] *Id.*

[167] *Id.* These memory 1 strategies are highly counterintuitive, because a player's whole history logically seems relevant to the question of defecting Later, Plotkin and his lab were able to prove longer memory strategies were more effect *See infra* Part II B

outcome of the previous interaction.[168] Using information from the previous interaction, "Tit for Tat" always makes the same move, whereas ZD strategies instead calculate probabilities for each of the four different outcomes of the Prisoner's Dilemma, and then play accordingly.[169] ZD strategies' approach guarantees a "linear relationship" between the two players' scores; they will not always achieve maximum points, but a ZD player can determine the other player's score "regardless of how [the other player] plays."[170] "The mathematical surprise offered up by Press and Dyson" thus concerned "expected payoffs" of the ZD strategies—namely, they showed using ZD strategies that "it is possible to force the [other player's] expected payoff"—an outcome that many within the world of game theory initial perceived as "mischief" because it would lead to domination of the other player, not cooperation.[171]

ZD strategies were initially viewed as a hostile or coercive class of strategies because of these strategies' ability to defeat other players regardless of their strategies. In their initial article, Press and Dyson described ZD strategies as "extortionate" where if one player alone, X, is using ZD strategies, she has the ability to "fix[] her strategy" and the other player, Y, "can maximize his own score only by giving X even more; there is no benefit in him defecting."[172] Early scholars looking to Press and Dyson focused on the extortionate aspects of ZD strategies, running computer experiments with humans where extortionate ZD strategies outperformed real players.[173]

These early game theory adopters of ZD strategies thus focused on the downsides these strategies produced, and argued that a player using a competitive strategy "never obtains less than the co-player" and extended ZD strategies from "memory one" games to longer-term memory games.[174] For example, researchers showed a mathematical tool

---

[168] *See* Christoph Adami & Arend Hintze, *Evolutionary Instability of Zero-determinant Strategies Demonstrates that Winning is Not Everything*, 4 NATURE COMM Aug 2013 at 2  According to this article, stochastic strategies were known even as early as the 1990s, though the article acknowledges Press and Dyson did revolutionize the field of game theory  This Article observes that had Axelrod paid more attention to the program DOWNING, which he acknowledged had the right impulse, he could have saved the field of game theory years, as DOWNING played stochastically

[169] *Id.*

[170] *Id.*

[171] *Id.* at 6  *See also* Stewart & Plotkin, *supra* note 17, at 10135

[172] Press & Dyson, *supra* note 162, at 10412

[173] Zhijian Wang, Yanran Zhou, Jaimie W  Lien, Jie Zheng & Bin Xu, *Extortion Can Outperform Generosity in the Iterated Prisoner's Dilemma*, 7 NATURE COMM  April 2016 at 2

[174] See Christian Hilber, Arne Traulsen, Karl Sigmund, *Partners or rivals? Strategies for the iterated prisoner's dilemma*, 92 Games & Economic Behavior 41 (2015); s*ee also* Alex McAvoy & Christoph Hauert, *Autocratic Strategies for Alternating Games*, 113 THEORETICAL POPULATION BIOLOGY 13, 20–21 (2017)

## *GAME THEORY*

known as a Markov chain led to new answers to other age old dilemmas used by the Chicago School and other game theorists with 2 x 2 payoff matrices such as Chicken and Snowdrift (discussed further *infra*), focusing on the "extortionate" aspects of these strategies.[175] Other scholars explored the dangerous potential of ZD strategies "master-slave" dynamics in which "slave" computer programs "cooperated" with each other in tournaments to intentionally lose to the "master" program intended to win the competition, becoming a means of exploitation.[176]

However, the exploitation insight eventually yielded to a more profound (and hopeful) answer. Press and Dyson noted in passing in their initial article that "if both X and Y are both witting of ZD," namely they both use ZD strategies, "then they may choose to negotiate to set the other's score to the maximum cooperative value" and "[u]nlike in naïve [Prisoner's Dilemma], there is no advantage in defection because neither can affect his or her own score and each can punish any irrational defection by the other."[177] And while the ZD insight started in computer

---

[175] Victor V Vasconcelos, Fernando P Santos, Francisco C Santos, and Jorge M Pacheco, *Stochastic Dynamics through Hierarchically Embedded Markov Chains*, 118 PHYSICAL REV LETTERS 058301-1, 058301-1–2 (2017) (obtaining new answers to "the snowdrift game in physics and economics, the hawk-dove game in evolutionary biology, and the chicken game in other contexts" and explaining the fact that "the evolution of cooperation" as well as "flocking behavior, voter dynamics, disease spread, diffusion of innovation, consensus formation, and peer influence" are all "complex time-dependent processes" that are "nonlinear" and that requires "defining a minimal (embedded) Markov chain whose solution estimates the limiting stationary distribution of the population and employing "a stochastic imitation process inspired in the Fermi distribution of statistical physics"); Jin-Li Guo, *Zero-determinant strategies in iterated multi-strategy games*, ARXIV:1409 1786v2 (2014) (discussing extortionate ZD strategies for Snowdrift); Lars Roemheld, *Evolutionary Extortion and Mischief Zero Determinant strategies in iterated 2x2 game*, HEIDELBERG UNIVERSITY ALFRED-WEBER-INSTITUTE FOR ECONOMICS BACHELOR'S THESIS (2013) (discussing extortionate ZD responses to 2x2 games including Chicken and Battle of the Sexes)

[176] TszChiu Au & Dana Nau, *Accident or Intention: That Is the Question (in the Noisy Iterated Prisoner's Dilemma)*, 5TH INTERNATIONAL JOINT CONFERENCE ON AUTONOMOUS AGENTS AND MULTIAGENT SYSTEMS May 2006, n 1, http://citeseerx ist psu edu/viewdoc/download?doi=10 1 1 61 9311&rep=rep1&type=pdf

[177] Press & Dyson, *supra* note, at 10412 Although noting that there is the option to cooperate when both players have a "theory of mind," Press and Dyson pessimistically concluded that extortion can prevail even when this is the case Even though Player X has the option to cooperate with Player Y, it is not required *Id.* Player X can still dominate Player Y unfairly, and Y's only alternative "to accepting positive, but meager, rewards is to refuse them, hurting both himself and X He does this in the hope that X will eventually reduce her extortion factor However [if X] has gone to lunch, then [Y's] resistance is futile " *Id.* The paper concludes "it is worth contemplating that, although an evolutionary player Y is so easily beaten within the confines of the IPD game, it is exactly evolution, on the hugely larger canvas of DNA based life, that ultimately has produced X, the player with the mind " *Id.* However, other scholars, most notably Plotkin's lab from the University of Pennsylvania, have shown that extortionate strategies do not maximize points overall, and that extortion eventually yields to cooperation, precisely because it is in the player's mutual self-interest to do so, as discussed *supra*

## *GAME THEORY*

science and physics,[178] it migrated into evolutionary biology. Alexander Stewart and Joshua Plotkin built on the extortionate ZD insight to show that, under the correct conditions, the correct or winning strategy to the Prisoner's Dilemma involves maximizing cooperation, not coercion and exploitation.[179] Stewart and Plotkin focused on the fact that ZD strategies involve players "with a theory of mind (i.e., a player who realizes that her behavior can influence her opponent's strategies)."[180] But, if both players have a "theory of the mind," "then each will initially try to extort the other, resulting in a low payoff for both."[181] In other words, in a computer tournament, each program must be able to understand the other along its repeated interactions. When these conditions exist, the "rational thing" is to "negotiate a fair cooperation strategy"—something along the lines of "Tit for Tat."[182] But ZD players have an "even better option":

> [B]oth can agree to unilaterally set the other's score to an agreed value (presumably the maximum possible). Neither player can then improve his or her score by violating this treaty, and each is punished for any purely malicious violation.[183]

Thus, extortion strategies of ZD players "work best when other players do not realize they are being extorted," but mathematically "ZD strategies that foster cooperation . . . receive the highest total payoff."[184] Stewart and Plotkin's work went further and re-ran Axelrod's old tournament with the addition of ZD strategies to see which would win. Their findings were unequivocal: They found that ZD strategies that fostered

---

[178] One physicist who reviewed this paper commented that the insight did not seem to properly belong to physics because the results appeared to be accomplished using linear algebra *But see* A Engle & A Feigel, *Single equalizer strategy with no information transfer for conflict escalation*, 98E PHYSICAL REVIEW 012415 (2018) (paper by two physicists in a physics journal analyzing ZD strategies in three and four dimensions); Zhi-Xi Wu & Zhihai Rong, *Boosting cooperation by in extortion in spatial prisoner's dilemma games*, E90 PHYSICAL REVIEW 062102, 062102-6 (2014) (physics paper studying the "spatial evolutionary prisoner's dilemma game with and without extortion by adopting the aspiration-driven strategy updating rule" and "[u]sing Monte Carlo simulations and generalized mean field approximations"); Xiongrui Xu, Zhihai Rong, Zhi-Xi Wu, Tao Zhou, and Chi Kong Tse, *Extortion provides alternative routes to the evolution of cooperation in structured populations*, E95 PHYSICAL REVIEW 05302 (2017) (physics paper studying ZD strategies using spatial distributions)
[179] Stewart & Plotkin, *supra* note 17, at 10135
[180] *Id.*
[181] *Id.*
[182] *Id.*
[183] *Id.*
[184] *Id.*

*GAME THEORY*

cooperation had an even higher payoff than "Tit for Tat," even though extortion strategies had the highest total win record in head to head one-on-one interactions (such as the strategy "always defect").[185] In particular, the strategy that received the highest score in the tournament that they called "ZDGTFT-2," which "force[d] the relationship $S_x - P = 2(S_y - P)$ between the two players' scores"—"offering a higher portion of payoffs above P"—"received the highest total payoff, higher even that Tit for Tat and Generous-Tit-for-Tat, the traditional players."[186] Thus, Stewart and Plotkin refined Dyson and Press' idea to confirm the instinct shared by certain early game theorists to act prosocially towards other players in a pact leading to pure cooperation. The outcome for the Prisoner's Dilemma as depicted by Plotkin? A tournament ranking showing the most generous ZD strategies indeed were the highest performers in the re-run Axelrod round-robin. But they lost out heavily in head-to-head matches, coming in with a score of zero and beaten by the program "AllD" or "Always Defect."

---

[185] *Id.*
[186] *Id.* (citing text from Figure 1 and noting that an extortion strategy "received a lower total payoff (because its opponents are not evolving), but it won more head-to-head matches than any other strategy, save Always Defect")

## GAME THEORY

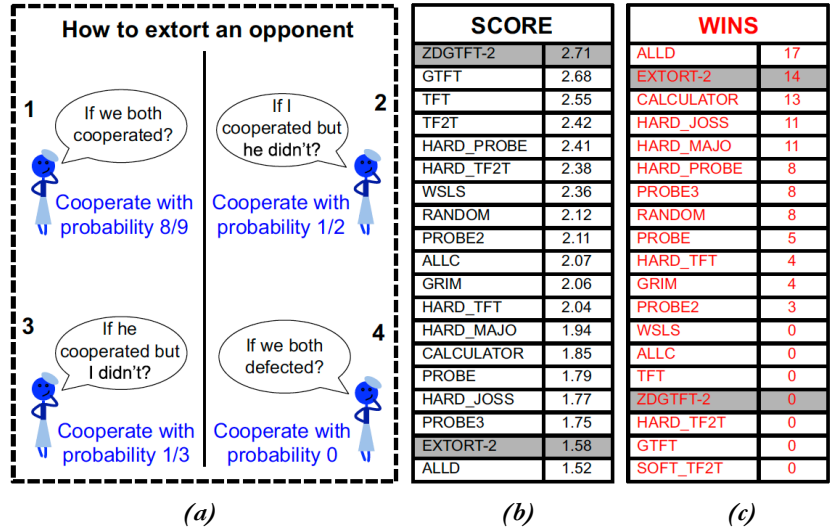

*Figure 3: Strategies ranked according to performance in a round-robin tournament of the iterated Prisoner's Dilemma.* "*A ZD strategy is specified in terms of four probabilities: the chance that a player will cooperate, given the four possibilities for both players' actions in the previous round.*" ***a)*** *Illustration of the ZDGTFT-2 strategy.* ***b)*** *The average score achieved by various strategies across all matches played.* "*A specific example, called Extort-2, forces the relationship Sx − P = 2(Sy − P) between the two players' scores. Extort-2 guarantees player X twice the share of payoffs above P, compared with those received by her opponent Y. A related ZD strategy that we call ZDGTFT-2 forces the relationship Sx− R = 2(Sy − R) between the players' scores. ZDGTFT-2 is more generous than Extort-2, offering Y a higher portion of payoffs above P.* ***c)*** *Within a match between two strategies, a strategy wins if it earns more rewards than its opponent. Notably ZDGTFT-2 wins no matches, yet achieves the maximum score among all games. Similarly, the AllD strategy wins every match but achieves the minimum total score. (Fig reproduced from Alexander J. Stewart & Joshua B. Plotkin,* Extortion and Cooperation in the Prisoner's Dilemma*, 26* PROC. NAT'L ACAD. SCI. U. S. AM. (PNAS) *10134 (2012)).*

Stewart and Plotkin's research was confirmed by other population biologists who showed that extortionate ZD strategies ultimately paved the way for systems based on cooperation. In dialogue with the Plotkin lab, Christian Hilbe, Martin Nowak, and Arne Traulsen wrote a critical paper that showed that, in large populations, "extortion plays a transient role" and ultimately paves the way for cooperation. In the case of ZD strategies, this "is the most abundant in large populations."[187]

---

*GAME THEORY*

These scholars concluded that "extortioners play a more prominent role in finite populations, where pairwise payoff advantages have a stronger effect. This is most intuitive when the population only consists of two individuals; since extortioners outperform their direct co-player by definition, extortion is expected to spread," but this expectation does not hold true for larger groups, where generosity prevails.[188] Indeed, extortioner scan "act as a catalyst for cooperation even while "apply[ing] a fully selfish strategy" by helping the population to escape from states with low payoffs."[189]

Hilbe's team also demonstrated that the results of ZD breakthrough were not "restricted to the case of the [P]risoner's [D]ilemma[] but c[ould] be extended to other social dilemmas."[190] Specifically, Hilbe's team, along with other scholars from Europe and China, showed canonical games such as the Chicken and Snowdrift yield different results when played in a pool of players as opposed to a two-player game.[191] In the Chicken game, two cars in opposite directions face each other and one must swerve to avoid a crash, with the one who serves being called "chicken." Chicken is used by the Chicago School to evaluate answers to many large-scale social questions.[192] According to Chicago School scholars, there is no stable equilibrium to this game in the prototypical payoff

---

[188] *Id.* at e77886

[189] *Id.* at e77886

[190] *Id.*

[191] *Id.* at e77886 (noting that in a snowdrift game, as in the prisoner's dilemma," "any sufficiently large initial population that yields a payoff less than S against itself can be replaced by more cooperative mutant strategies with higher baseline payoffs      which can only be left by neutral invasion of altruists ")

[192] Christopher S Yoo, *Beyond Coase: Emerging Technologies and Property Theory*, 160 U PENN L REV 2190, 2217-19 (2012) (citing Anatol Rapoport & Albert M Chammah, Th*e Game of Chicken*, Am Behav S, Nov 1966, at 10); Lee Anne Fennell, *Common Interest Tragedies,* 98 NW U L REV 907, 941-48 (2004); ROBERT SUGDEN, THE ECONOMICS OF RIGHTS, CO-OPERATION AND WELFARE 58-62, 128-32 (1986))

## *GAME THEORY*

matrix depicted below, with the players alternating outcomes as shown by the diagram.[193]

**Chicken Game**

|  | Player 2 Cooperate | Player 2 Defect |
|---|---|---|
| Player 1 Cooperate | 3, 3 | 2, 4 |
| Player 1 Defect | 4, 2 | 1, 1 |

*Figure 4: Payoff matrix for the chicken game, a commonly used model for conflict scenarios. The classical view is that players are better off taking the opposite choice of their opponent, resulting in a lack of a stable equilibrium. (Figure reproduced from* Christopher S. Yoo, *Beyond Coase: Emerging Technologies and Property Theory*, 160 U. PENN. L. REV. 2190, 2217-19 (2012)).

Unlike the Prisoner's Dilemma, where there is a lack of communication between players, Chicken encourages "bluffing," where one player tries to convince the other player that she is "suicidal" and "irrational" enough to be willing to cause a crash, thus inducing the other car to swerve.[194] As such, the game is considered even more adversarial than the Prisoner's Dilemma[195] because "one party willfully *creates* a conflict by challenging the other and threatens to *destroy* an already enjoyed common interest; the defending party may reciprocate with a similar threat" and compromise only occurs when each views the other side as wholly irrational.[196] In the Chicago School, Chicken is used as a game theoretic model of international conflicts where parties are "willing to stand firm only if one is fairly confident that the opponent will back down."[197] Chicken is also used as a model for the "tragedy of the commons," with theorists explaining how the incentives for players are to "maneuver for larger shares of the surplus" through engaging in "holdout" behavior. Scholars note that even though it was not the accepted answer, "mutual cooperation" is the obvious solution.[198] And, as in the Prisoner's

---

[193] *Id.*
[194] *Id.*
[195] *Id.*
[196] Glenn H Snyder, *"Prisoner's Dilemma" and "Chicken" Models in International Politics*, 15 INTERNATIONAL STUDIES QUARTERLY 66, 84 (1971)
[197] *Id.* at 93 (noting that Chicken was not a good model for the Cuban missile crisis and also that it would not be a good model for nuclear war given the possibility of escalation)
[198] Fennell, *supra* note 192, at 944

*GAME THEORY*

Dilemma, where Tit for Tat replaces the Pure Nash Equilibrium of "Defect, Defect," it is accepted within the purview of the Chicago School that playing Chicken repeatedly increases the chances of cooperation, and that over the long run forgiving strategies fair better.[199] Similarly, the Snowdrift Game, a variation on Chicken, models who will shovel a road when it is covered with snow when "two drivers are caught in a blizzard and trapped on either side of a snowdrift. They can only get home if the snow is shoveled. Thus[,] they can either choose shoveling (cooperation, C) or staying in the car (defection, D)."[200] The classical approach to Snowdrift, like the approach to Chicken, takes the view that "the best choice of the driver always depends upon her opponent: it means if her opponent's strategy is [cooperate], it is better to choose [defect] and vice versa."[201]

      Generous ZD strategists in Europe and in China demonstrated that these classical answers to Chicken and Snowdrift are outdated when these games are played evolutionarily in a population of players. In a multiplayer Snowdrift Dilemma (where instead of two individuals who need to clear the snow, there are groups), both extortionate as well as generous ZD strategies dominate.[202] Crucially, for generous ZD strategies a key variable in the Snowdrift Dilemma is the threshold number of cooperators, which is depicted in the Figure 5 below on the left in blue in an 8-player game, with a cooperator threshold.

---

[199] Yoo, *supra* note 192, at 2219 (citing Bengt Carlsson & K Ingemar Jönsson, *Differences Between the Iterated Prisoner's Dilemma and the Chicken Game Under Noisy Conditions*, 2002 PROC ACM SYMP ON APPLIED COMPUTING 42, 46 ("With increased noise forgiving strategies become more and more successful in [an iterated Chicken Game] while repeating and revenging strategies are more successful in [an iterated Prisoner's Dilemma] ")
[200] Qian Zhao, Chuyi Guo, Zhihai Rong, *The Evolution of Submissive Clusters in the Spatial Snowdrift Game*, 3RD INTERNATIONAL SYMPOSIUM ON AUTONOMOUS SYSTEMS (ISAS) (2019)
[201] *Id.*
[202] K Frieswijk, A Govaert, M Cao, *Exerting Control in Repeated Social Dilemmas with Threshold*, 53 IFAC PAPERS ONLINE 16946 (2020)

*GAME THEORY*

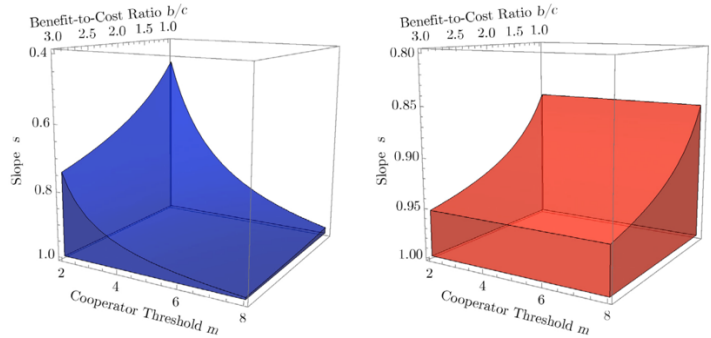

*Figure 5: "Regions of strategy-existence in an 8-player snowdrift dilemma with a cooperator threshold m, for: generous strategies (left); extortionate strategies (right)." Notably, the payoff of the generous strategy (left) depend also on the number of cooperators whereas the extortionate strategy (right) does not. (Fig reproduced from K. Frieswijk, A. Govaert, M. Cao,* Exerting Control in Repeated Social Dilemmas with Threshold*, 53 IFAC PAPERS ONLINE 16946 (2020).*

Other scholars, repeating the game over time in a mixed population of cooperators and defectors, have shown that populations of cooperators form clusters together that displace the defectors. [203] As depicted in Figure 3 below, populations of cooperators (referred to as sZD players depicted in blue) gradually replace the defectors (depicted in red) in the square lattice population.[204]

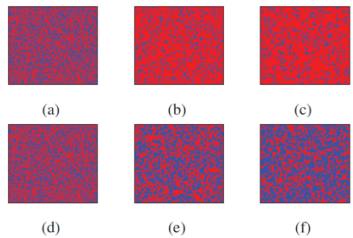

*Figure 6: Simulations of iterated multiplayer Snowdrift Dilemma from an initial state to the tenth iteration to the final generation. Blue dots indicate sZD players (cooperators) and red dots indicate defectors. In panels (a)-(c), as time evolves, "the sZD players will form the filament-like clusters . . . when they meet defectors in the square lattice [and will not prevail in payoffs.] On the contrary, in (d)-(f), the submissive individuals will get into a large cluster to defend the invasion of defectors." (Figure reproduced from Qian Zhao, Chuyi Guo, Zhihai Rong,* The Evolution of Submissive Clusters in the Spatial Snowdrift Game*, 3RD INTERNATIONAL SYMPOSIUM ON AUTONOMOUS SYSTEMS (ISAS) (2019).)*

---

[203] Zhao, Guo, & Rong, *supra* note 200, at 156
[204] *Id.*

## *GAME THEORY*

However, even though the bulk of the existing literature focuses on population games and new answers to the Prisoner's Dilemma, Chicken, and Snowdrift, ZD strategies can yield new answers to 2 x 2 games played in and of themselves. This disproves the Chicago School on its own terms. Lars Roemheld proved that in 2 x 2 symmetric two player games played infinitely, "a mischievous X may unilaterally force $\pi_y$ to any value between Y's pure strategy maximin payoff (r = max{S; P}) and the lower of his two highest payoff (s:= min{R; T})."[205] The extortionate ZD equilibriums from this thesis for Chicken are depicted below.

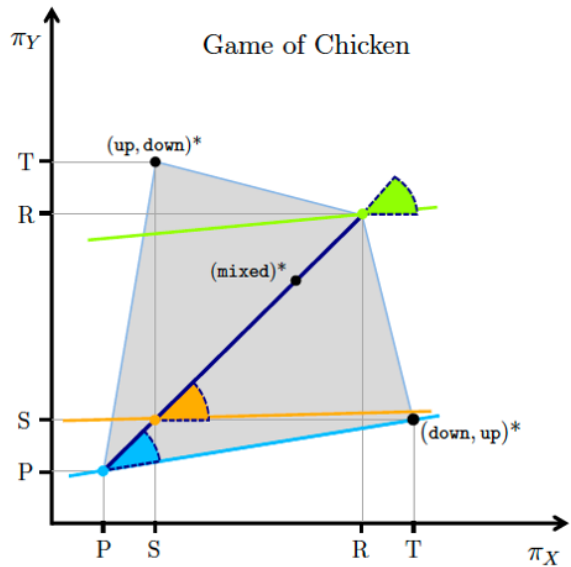

*Figure 7: ZD extortion strategies. The green tip of at the corner of the feasible region provides a cooperative equilibrium, contrary to the theory of the Chicago School. The blue and orange regions represent other equilibriums, namely "the lower edge of the set of feasible payoffs (the line (down, down) → (down, up), the same one which could be obtained by simply always playing down. S, T refer to outcomes of the payoff matrix as in Figure 1. (Figure reproduced from Lars Roemheld,* Evolutionary Extortion and Mischief Zero Determinant strategies in iterated 2x2 game, HEIDELBERG UNIVERSITY ALFRED-WEBER-INSTITUTE FOR ECONOMICS BACHELOR'S THESIS (2013))*.

---

[205] *See* Roemheld, *supra* note 175 at 20-33 (noting there was no extortionate ZD strategy for the 2 x 2 game Battle of the Sexes, and discussing at length, as in the Press and Dyson paper, that extortionate strategies can result in an ultimatum game that leads to exploitation and interpreting these results pessimistically); *see generally* Guo, *supra* note 175 (confirming extortionate ZD strategies produce new answers for Chicken in 2 x 2 games using Markov matrices, though not depicting two extortionate strategies playing against one and other)

## *GAME THEORY*

As can be seen from the figure above, at the equilibrium at the "upper limit" of two ZD extortionate strategies—i.e., the green region in the diagram—is one where both players have the option to maximize cooperation by choosing "up," or "always swerve." This answer completely contradicts the theory of Chicken in the Chicago School, which is that there is no equilibrium (or, that the equilibrium is "up, down" or "swerve, crash") to the necessarily adversarial nature of the game. Thus, this work unwittingly confirms that when both extortionate players are "witting" of ZD and have a "theory of mind" both players can choose pure cooperation to maximize points instead of being adversaries.[206] Thus, what little research there is on two player games suggests that ZD strategies can produce different, and better, answers than the Chicago School to classical games so long as they are played repeatedly in infinite games.

Finally, the behavior exhibited by generous ZD strategies is present not only in math and physics but is also in nature. This shows that it is possible to have functioning and stable societies based on the pure cooperation insight even though many species are exempt. Evolutionary biologists have examined altruistic behavior in eusocial insects like ants and wasps, where workers sacrifice their own wellbeing for the good of the colony and for the "leader" due to the genetic fitness benefit that accrues from this behavior[207] and spiteful bacteria that poison

---

[206] In my original email correspondence with Professor Oechssler who supervised Lars Roemheld he indicated that these results "can be applied (in [their] original form) to all 2 x 2 games" but that equilibriums results would differ depending on the type of ZD strategy used, namely "extortive, fair, [or] altruistic" See Email from Joerg Oechssler, Alfred-Weber Institute for Economics, to Cortelyou C Kenney, Academic Fellow, Cornell Law School (Nov 11, 2020) (on file with author)  However, contrary to the clear results of depicted in the graph above, Professor Oechssler incorrectly suggested that "always swerve" was not an equilibrium for Chicken because it was "inefficient"; incorrectly stated that ZD strategies are  "pretty mechanical" and do not involve a "theory of mind;" incorrectly stated that Tit for Tat is the "optimum" result for the Prisoner's Dilemma; and incorrectly implied research from Hilbe's team showed ZD strategies were  "no solution to social dilemmas" even though the paper itself explicitly shows with proper enforcement they can be after the initial failure of cooperation  *Id; see infra* Part II B  While experts are experts for a reason, ultimately allowing the underlying research to speak for itself is the best approach to avoid unwittingly dismissing new, and potentially very important, ideas *See* Press & Dyson, *supra* note, at 10412  As can be seen from the diagram above, and from Press and Dyson's paper, pure cooperation is not required and there are other equilibriums that reduce to an ultimatum game, though that is not in the mutual best interest of the players  *But see* Sau-Him Paul Lau & Vai-Lam Mui, *Using turn taking to achieve intertemporal cooperation and symmetry in infinitely repeated 2 x 2 games,* 72 THEORY 167, 182 (2012) (economics journal article advocating for cooperation in Chicken, but falling short of maximal cooperation achieved by ZD strategies)

[207] JAMES H  HUNT, THE EVOLUTION OF SOCIAL WASPS (2007)  Both altruism and spite are, in the Chicago School and in classical game theory, considered carve-outs to the Prisoner's Dilemma, which focuses exclusively on selfish behavior (though Judge Posner explains that, in his view, biological impulses such as the desire for revenge are rational)  *See, e.g.,* Picker, *supra* note 4, at 4 (also ignoring reputational consequences in the Prisoner's Dilemma for snitching)

*GAME THEORY*

competitors even when this act also harms the bacteria themselves—neither of which the ZD models account for.[208] But the ZD insight of pure cooperation can be seen in species that engage in reciprocal altruism for non-genetic reasons. One example is vampire bats, whose community members regurgitate blood to individuals who have not fed the night before even though these individuals are not closely connected by kinship.[209] In exchange, vampire bats feed other bats that stay in the roost, babysit, and protect young other than their own from predators.[210]

Evolutionary biologists have also looked to other species acting cooperatively in the behavior of monarch butterflies. Monarchs make a long, multi-generational migration every year from Canada to Mexico, and to protect both themselves and each other from cold temperatures the butterflies cluster together in large formations to prevent themselves from freezing.[211] This clustering behavior is a paradigmatic example of how communities are "all in it together" and how benefiting the individual benefits the collective and vice-versa—whether the behavior is due to some benefit to the individual (in the case of ants and wasps) or to the community as a whole (in the case of vampire bats).

---

[208] *See, e.g.*, R Fredrik Inglis et al , *Spite and Virulence in the Bacterium* Pseudomonas Aeruginosa, 106 PROC NAT'L ACAD SCI U S AM (PNAS) 5703, 5703 (2009)
[209] *See, e.g* , Gerald G Carter & Gerald S Wilkinson, *Food Sharing in Vampire Bats: Reciprocal Help Predicts Donations More Than Relatedness or Harassment*, 280 PROC ROYAL SOC'Y B 1 (2013); Gerald S Wilkinson et al , *Non-kin Cooperation in Bats*, 371 PHIL TRANSITIONS B September 2016 at 1
[210] Wilkinson et al , *supra* note 209, at 1-3
[211] Lincoln P Brower et al , *Monarch Butterfly Clusters Provide Microclimatic Advantages During the Overwintering Season in Mexico*, 62 J LEPIDOPTERISTS' SOC'Y, 177, 177, 185-88 (2008)

*GAME THEORY*

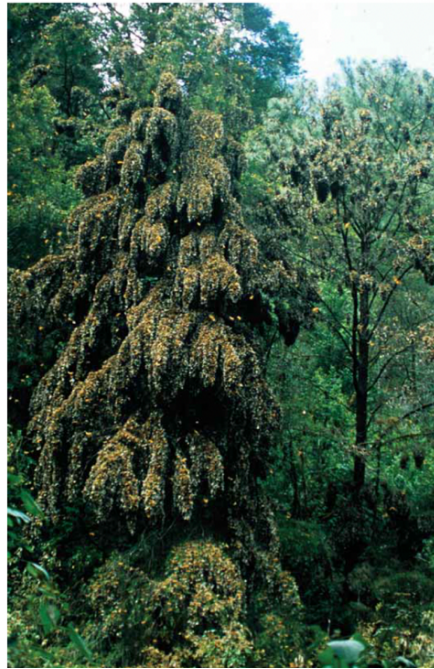

*Figure 8: Monarchs exhibiting collective behavior by clustering for warmth. "The density of clustering monarchs varies according to the foliage architecture of the tree species on which they settle. Note the exceedingly dense clusters on the oyamel fir (left foreground) and the much smaller ball-like clusters on the pine (right background). Photo taken in the Ojo de Agua ravine on Cerro Pelon in the state of Mexico, 13 Feb 2004." (Figure reproduced from Lincoln P. Brower et al., Monarch Butterfly Clusters Provide Microclimatic Advantages During the Overwintering Season in Mexico, 62 J. LEPIDOPTERISTS' SOC'Y, 177, 177, 185-88 (2008).)*

B. Creating, Maintaining, and Scaling Cooperation

The pure cooperation insight in turn leads to the natural (and still unanswered) question of how the conditions that enable pure cooperation can be created, maintained, and scaled. This question vexed Stewart and Plotkin and other scholars who found that in communities based on two player games, pure cooperation is not stable and can collapse "when the ratio of costs to benefits becomes too high."[212] Shortly after their

---

[212] Alexander J. Stewart, *New Take on Game Theory Offers Clues on Why We Cooperate*, Conversation (Mar 11, 2015, 12:02 PM EDT), https://theconversation.com/new-take-on-game-theory-offers-clues-on-why-we-cooperate-38130. Plotkin's lab also recently came out with a paper that shows—as our country is currently facing an extreme amount of polarization—"that if the environment becomes sufficiently risk adverse, only the high-politzerization equilibrium is stable." *See* Alexander J. Stewart, Joshua Plotkin, & Nolan McCarthy, *Inequality, Identity, and Partisanship: How Redistribution Can Stem The Tide of Mass Polarization*, 118 PROC. NAT'L ACAD. SCI. U.S. AM. 1, 3 (PNAS) (2021). According to the article, "[t]he

*GAME THEORY*

pathbreaking and hopeful work showing that pure cooperation can be achieved through mutual pact based on reciprocity and generosity, Stewart and Plotkin published a second, much more pessimistic, paper. It showed that "as individuals maximize the benefits of cooperation," they often pave the way for its collapse.[213] The key was a phenomenon known as "coevolution"—species evolving in relation to one another with biofeedback between the species and their environment.[214] Again using the Iterated Prisoner's Dilemma, Stewart and Plotkin found that:

> [W]hen both strategies and payoffs coevolve[,] there is a striking reversal of fortunes. Evolution favors increasing the benefits of mutual cooperation as well as the benefits of unilateral defection. . . Paradoxically, defection comes to dominate even as the payoffs for mutual cooperation continually increase. Moreover, this collapse of cooperation is often accompanied by an erosion of mean population fitness.[215]

The paper concluded that "cooperation will always collapse when there are diminishing returns for mutual cooperation."[216] As Stewart explained: "Suppose everyone in the team really does go the extra mile when they work on a project. Then every member of the team knows he or she has relatively little to lose by slacking off, because everyone's extra effort will still carry them."[217] This "paradox," Stewart explained, is "exactly what we see in evolving games—cooperating players contribute ever greater effort to cooperation, only to make it easier for defectors to take hold," or, "the more we cooperate, the less likely others are to do the same."[218] This finding was replicated by Hilbe's team in 2015, which found

---

only remedy for reversing a polarized state, under our analysis, requires either a shock [] or a sufficiently good economic environment coupled with collective action by a population who change strategies simultaneously " *Id.*

[213] Alexander J Stewart & Joshua B Plotkin, *Collapse of Cooperation in Evolving Games*, 111 PROC NAT'L ACAD SCI U S AM (PNAS) 49, 17558–17563 (2014)

[214] *Id.*

[215] *Id.*

[216] *Id.*

[217] Alexander J Stewart, *New Take on Game Theory Offers Clues on Why We Cooperate*, Conversation (Mar 11, 2015, 12:02 PM EDT), https://theconversation com/new-take-on-game-theory-offers-clues-on-why-we-cooperate-38130

[218] *Id. See also* Erol Akçay, *Collapse and rescue of cooperation in evolving dynamic networks*, 9 NATURE COMM 2692 (2018)

## *GAME THEORY*

cooperation eventually collapsed through the entrance of invaders: In their words, while "extortionate strategies can act as a catalyst for cooperation, as they are able to subvert a population of defectors irrespective of the size of the group."[219] Although cooperation can collapse with the influx of invaders, Hilbe's team identified strategies for effective enforcement of cooperation. These strategies ensure that once cooperation has been established, it can be maintained with the help of effective enforcement mechanisms such as "the formation of alliances or institutions, or additional pairwise interactions between group members. "[220]

And so, as in real life, there are some teams that simply defy the odds and escape the paradox with everyone working at incredibly high levels of performance and team cohesion—think NASA's moonshot or hospitals like Massachusetts General noted for standout patient outcomes. These are teams where the community itself is defined by a culture of mutual accountability and excellence.[221] Perhaps understanding this, Stewart and Plotkin ended their paper with a caveat, holding out hope that the initial pure cooperation insight was not so limited: "Decoupling strategy evolution from payoff evolution may not be appropriate in all biological contexts. Alternative modeling frameworks such as continuous games, which allow players to modulate their levels of investment in a social interaction can provide a contrasting or complementary perspective on the evolution of cooperation."[222] They concluded that "the predicted prevalence of [cooperation and defection] depends critically on the payoffs resulting from social interactions. Understanding the feedback between strategy evolution and payoff evolution is therefore critical for understanding social interactions in natural populations."[223] The paper also showed that after the collapse of cooperation, based on certain constraints, "cooperative strategies recover to high frequencies" and "following the collapse of cooperation, the Prisoner's Dilemma is

---

[219] Christian Hilbe, Bin Wu, Arne Traulsen, *Evolutionary performance of zero-determinant strategies in multi-player games*, 374 J THEOR BIOL 115–124 (2015)

[220] *Id.*

[221] *See* Atul Gawande, *Amid the Coronavirus Crisis, a Regimen for Reëntry*, THE NEW YORKER (May 13, 2020), https://www newyorker com/science/medical-dispatch/amid-the-coronavirus-crisis-a-regimen-for-reentry (describing the culture of Massachusetts General during the pandemic)

[222] Stewart and Plotkin also noted that the existence of "both cooperators and defectors" such as the "marine bacteria *Vibrionaceae*" are "often found at appreciable frequencies in nature " Stewart and Plotkin, *supra* note 213, at 17561

[223] *Id.*

*GAME THEORY*

replaced" and "self-cooperating strategies recover to high frequencies in populations."[224]

In 2015, Stewart and Plotkin predicted that it would be "a long way off" before their evolutionary biology research would be useful to understand human cooperation and suggested scientists should spend their efforts concentrating on micro-organisms.[225] But new innovations in their work soon proved otherwise. First, a 2016 paper suggested that that cooperation might first be tackled by varying the strategy in smaller communities by using strategies with longer memories (ZD strategies rely on a memory of 1). These longer memory scenarios can explain how cooperation evolved even when the benefits to cooperating are low because players are better able to defend against invaders.[226] Another 2016 found that concerns of cooperation collapsing could be mitigated through behavioral diversity.[227] Moving beyond the Prisoner's Dilemma to multiplayer games with "nontransitive payoff structures, under which no hierarchical ordering of payoffs is possible," could help a diverse population resist invaders for the "mutual benefit of the population."[228] In such scenarios, cooperation dominates where individuals are not faced with all-or-nothing choices and where contributions can be made slowly over time.[229]

Based on the foregoing, the generous ZD community eventually concluded that cooperation works better in multiplayer rather than two-player games over the long run. In other words, cooperation attained through generous ZD strategies can collapse in two-player games such as the Prisoner's Dilemma when "evolution initially produces a rapid loss of cooperative strategies and an increase in the frequency of defecting strategies."[230] But ZD strategies allow cooperation to recover in larger populations playing multiplayer games among populations greater than

---

[224] *Id.* at 17561-62
[225] Alexander J Stewart & Joshua B Plotkin, *The Evolvability of Cooperation Under Local and Non-Local Mutations*, 6 GAMES 231, 243 (2015)
[226] Alexander J Stewart & Joshua Plotkin, *Small Groups and Long Memories Promote Cooperation*, 6 SCI REPS. June 2016, at 1
[227] Alexander J Stewart, Todd L Parsons & Joshua B Plotkin, *Evolutionary Consequences of Behavioral Diversity*, 113 PROC NAT'L ACAD SCI U S AM (PNAS) E7003, E7008 (2016)
[228] *Id.*
[229] *See also* Alexander J Stewart, Todd L Parsons & Joshua B Plotkin, *Evolutionary Consequences of Behavioral Diversity*, 113 PROC NAT'L ACAD SCI U S AM (PNAS) E7003, E7008 (2016) (analyzing non-binary choices through behavioral diversity)
[230] *See id.;* Alexander J Stewart & Joshua B Plotkin, *Collapse of Cooperation in Evolving Games*, 111 PROC NAT'L ACAD SCI U S AM (PNAS) 49 (2014)

## *GAME THEORY*

one hundred.[231] And Plotkin and his lab show that for multiplayer games over population of ten, generous ZD strategies can dominate and only "mildly punish" defectors, producing a surprising result: whereas "[o]ne might expect such strategies to be vulnerable to replacement by defector strategies, [Plotkin and his lab] have shown that the reverse is true."[232]

Given the potential for cooperation to collapse after it has been created, the obvious question to ask is how to sustain it. To answer this question, it is wise to provide a brief history of the age-old problem of the "tragedy of the commons" about resource distribution within a world of limited goods.[233] In studying enforcement mechanisms to sustain cooperation, Plotkin's lab built on the work of economists and sociologists from this period onward—including the Nobel Laureate Elinor Ostrom, the second woman ever to win the prize in her field whose work out of Indiana University focused on common pool resources grew out of her graduate study of groundwater basins.[234]

Ostrom's work famously examined how to promote cooperation by looking at local communities around the world to yield a series of factors that, in her view, were likely to increase the chances of functioning communities that successfully shared finite resources from forests to fish. Her seminal work examined indigenous communities as well as ancient societies around the world with populations of fewer than 15,000 people that remained relatively stable over time.[235] Ostrom extrapolated seven factors to explain why these communities were able to work together so well. She showed why they had been able to solve the intractable "free rider" problem without necessarily resorting to top-down or

---

[231] *Id.; see also* Christian Hilbe, Bin Wu, Arne Traulsen, & Martin A. Nowak, *Cooperation and Control in Multiplayer Social Dilemmas*, 111 PROC. NAT'L ACAD. SCI. U S AM. (PNAS) 16425, 16425 (2014); Christian Hilbe, Bin Wu, Arne Traulsen, Martin A. Nowak, *Evolutionary Performance of Zero-Determinant Strategies in Multiplayer Games*, 374 J. THEORETICAL BIOLOGY 115, 2 (2015) (noting that "evolution eventually stabilizes at either generous strategies or extortionate strategies, depending on two key parameters capturing the essence of ZD strategies")

[232] Alexander J. Stewart & Joshua B. Plotkin, *From extortion to generosity, evolution in the Iterated Prisoner's Dilemma*, 38 PROC. NAT'L ACAD. SCI. U S AM. (PNAS) 15348, 15352 (2013)

[233] Garrett Hardin, *The Tragedy of the Commons*, 162 SCIENCE 1243 (1968)

[234] *See* ELINOR OSTROM, GOVERNING THE COMMONS: THE EVOLUTION OF INSTITUTIONS FOR COLLECTIVE ACTION (1990); THE COMMONS IN THE NEW MILLENNIUM: CHALLENGES AND ADAPTATION (Nives Dolšak & Elinor Ostrom eds., 2003); *see also* MANCUR OLSON, THE LOGIC OF COLLECTIVE ACTION: PUBLIC GOODS AND THE THEORY OF GROUPS (1965)

Ostrom's work has also influenced legal scholars in the field of law and political economy like Amy Kapczynski, who authored a paper on a network for open science and influenza trackers—of "IP without IP"—where "the Network played [a role] in deliberately cultivating a sense of community, equality, and trust." *See* Amy Kapczynski, *Order without Intellectual Property Law: Open Science in Influenza*, 102 CORNELL L. REV. 1539, 1549, 1604-08 (2017)

[235] OSTROM, *supra* note 234, at 95

## GAME THEORY

state enforced solutions.[236] Later in her career, Ostrom revised her thesis and claimed that there were anywhere between twenty to thirty factors in these communities and warned against overly-simplistic models that seemed elegant on paper, but that in practice failed to explain the behavior of the underlying communities.[237] Toward the end of her career, Ostrom argued that the "key is assessing which variables at multiple tiers across the biophysical and social domains affect human behavior and social–ecological outcomes over time."[238]

But more recent work from Plotkin's lab, most notably Andrew Tilman's, suggested the contrary. There are ways to mathematically model local, bottom-up solutions using game theory or other mathematical games that are more complex than the Prisoner's Dilemma.[239] However, these more complex models still reflect the "simple" "intuition" of the ZD strategies mathematical models discussed above.[240] Specifically, they lead to the conclusion that there is feedback between the players and their environment, and that modeling this bidirectional feedback and the relationship between the players and their environment holds the key to solving various social problems.[241] Tilman and his co-authors focus on fisheries as one natural example of such feedbacks in their papers as models that potentially hold the key to other, broader social problems like global climate change and the field of deliberative decision-making, which then can "set[] the stage for the success of less costly, non-deliberative decision-making."[242]

In an early paper, Tilman along with Simon A. Levin and James R. Watson demonstrated that fisheries that employ globally around 37 million workers can serve as game theoretic models to potentially crack

---

[236]OSTROM, *supra* note 234, at 273-278

[237]*See* Elinor Ostrom, *A General Framework for Analyzing Sustainability of Social-Ecological Systems*, 325 SCIENCE 419, 421 (2009); Elinor Ostrom, *A Diagnostic Approach for Going Beyond Panaceas*, 104 PROC NAT'L ACAD SCI U S AM (PNAS) 15181 (2007) (hereinafter Ostrom, *Panaceas*)

[238]Ostrom, *Panaceas*, *supra* note 237, at 15183

[239]Andrew R Tilman, Joshua B Plotkin, & Erol Akçay, *Evolutionary Games with Environmental Feedbacks*, 11 NATURE COMM 1 (2020)

[240]*Id.* at 9  Note that many game theorists believe the intuitive answer to the Prisoners' Dilemma is "cooperate, cooperate," and have advocated for this answer, though not using Zero Determinant strategies  *See* McAdams, supra note 8, at 217-218 n 31 (collecting extensive on cooperate, cooperate all pre-2012, though noting that this answer is an "error" and that the "confess, confess" solution to the Prisoners' Dilemma is really "one of the coordination games discussed below ")

[241]*Id.* at 2

[242]*Id.* at 2 (citing David G Rand et al, *Cyclical Population Dynamics of Automatic Versus Controlled Processing: An Evolutionary Pendulum*, 124 PSYCH REV 626, 626 (2017))

## *GAME THEORY*

other important social problems.[243] Tilman's work examines the conditions within fisheries and how to create and sustain cooperation. He reaches two conclusions that suggest that cooperation may be simpler than Ostrom argues. First, he and his co-authors show that the goal of efficiency or resource maximization is not the best for the community.[244] Better was to pursue a "sub-optimum" catch level that enabled fishers not to overfish and to agree to limit their consumption in a sustainable way.[245] Second, Tilman and co-authors showed that to outperform top-down government control—especially in jurisdictions with underenforcement—fisheries needed to rely on social norms to enforce their own local agreements to fish at these "suboptimum" levels.[246] Fisheries could enforce these agreements through ostracism and social punishment for any harvesters who attempted to "free ride" and break the agreement to preserve equity and fairness for their members.[247] Ultimately, their solution enables "sustainable resource use" that, in the long run, benefits both the fish and the fishers.[248] Their insight, however, relied on the notion that fishers should not follow their "individual incentives" in the "Nash equilibrium level of effort"[249]—but in the end, the individual incentive of each fisher is *better* should that individual choose to participate in a binding social contract if the alternative is to be excised from the community.[250]

Tilman concluded that "bottom up management might be able to emerge even in communities where the strong social bonds needed to enforce social norms are not present."[251] The caveat to his work is that they did not examine "heterogeneity" among the communities they studied and they also postulated that "multiple (small) revenue-sharing clubs" may be more effective than a single large "common-pool resource

---

[243] Andrew R. Tilman, Simon Levin, & James R. Watson, *Revenue-Sharing Clubs Provide Economic Insurance and Incentives for Sustainability in Common-Pool Resource Systems*, 454 J. THEORETICAL BIOLOGY 205 (2018); *see also* Andrew R. Tilman, James R. Watson, & Simon Levin, *Maintaining Cooperation in Social-Ecological Systems: Effective Bottom-Up Management Often Requires Sub-Optimal Resource Use*, 10 J. THEORETICAL ECOLOGY 155 (2017)
[244] Tilman, Watson, & Levin, *supra* note 243, at 163
[245] *Id.* at 156
[246] *Id.*
[247] *Id.*
[248] *Id.* at 161, 153
[249] *Id.* at 162
[250] Tilman, Plotkin, & Akçay, *supra* note 239, at 4
[251] ANDREW ROBERT TILMAN, ECOLOGICAL, ECONOMIC AND SOCIAL MECHANISMS FOR COMMON-POOL RESOURCE MANAGEMENT 61 (September 2017) (Ph.D. dissertation, Princeton University) (on file with author)

## *GAME THEORY*

system."[252] The bottom line, however, is that "risk mitigation strategies can be used as a catalyst for common-pool resource[s]" in certain communities "to cooperatively self-govern, leading to both economic and environmental wins."[253]

Third, in terms of using social mechanisms to enforce non-compliance, there are plenty of examples in the real world where using the social fabric to inform other community members that their behavior is harmful is actually more effective, efficient, and humane than government regulation and punishment.[254] In a paper from 2020, Tilman, Plotkin, and Erol Akçay discussed the incentive to "follow the gold rush"—that is to cheat, or otherwise break with a cooperative system in order to benefit oneself.[255] They discuss how preventing this type of behavior is possible (though difficult) and can be achieved by providing the correct incentives. By using these incentives, eventually the population can achieve "bistability"— a stable mixture of both cooperation or exploitation ("gold rush" or "free rider" behavior).[256] Building on this research, as part of achieving stability in situations such as these, Hilbe's team noted players in ZD multiplayer games can form "alliances" and "once an alliance has [a] critical mass, there are no bounds for extortion."[257] Indeed, in this scenario, players who coordinate among themselves in groups "can trigger a positive group dynamic among outsiders" and that hold the key to solving widespread social dilemmas such as the public goods game.[258]

Once cooperation is established, one crucial factor to ensure it is maintained is the importance of reputation within the community as measured by "empathetic perspective taking" or scores measured by neutral third parties broadcast to the members of the community.[259] For

---

[252] Tilman, Levin, & Watson, *supra* note 243, at 213-14

[253] *Id.*

[254] Tilman, Watson, & Levin, *supra* note 243, at 163

[255] Tilman, Plotkin, & Akçay, *supra* note 239, at 4

[256] *Id.* at 5

[257] Hilbe, *Cooperation and Control*, *supra* note 231

[258] *Id.*

[259] Arunas L Radzvilavicius, Taylor A Kessinger, & Joshua B Plotkin, *Adherence to Public Institutions That Foster Cooperation*, 12 NATURE COMM June, 2021 at 1 Note there is a vast scholarship both in law and in sociology on the problematic nature of reputation *See generally* Yonathan Arbel, *Reputation Failure: The Limits of Market Discipline in Consumer Markets*, 54 WAKE FOREST L REV 1239, 1239 (2019) (arguing "reputational information is beset by participation, selection, and social desirability biases that systematically distort it" and that "these distortions are inherent to most systems of reputation and that they make reputation far less reliable than traditionally understood"); Marion Dumas, Jessica L Barker, and Eleanor A Power, *When does reputation lie? Dynamic feedbacks between costly signals, social capital and social*

## *GAME THEORY*

example, according to Radzviavicius, Kessinger, and Plotkin in a 2021 paper, "[i]nteractions in modern societies often involve cooperation with strangers, and so people must rely on reputation scores provided by institutions."[260]

According to Radzviavicius, Kessinger, and Plotkin, game theory models show populations can remain stable when people cooperate "only with players with good reputations," or, in the alternative, "empathy can reduce the rate of misunderstandings and unjustified defection," and "empathy can itself evolve through social contagion, inducing high rates of cooperation typical of societies that an established public monitoring system."[261] But, when institutions function, "moral assessment can spread naturally" and as mathematical models such as two-player games and Monte Carlo simulations show, "there are large basins of attraction toward cooperation when public reputations are provisioned by highly tolerant institutions."[262] Indeed, institutions in which "a single interaction with a bad individual can trigger a cascade of punishment and defection" will eventually "lead[] to low cooperation rates" and "[i]n these cases tolerate institutions help individuals being assigned a bad reputation from occasion interactions with bad players, and so a high frequency of cooperating discriminators is less likely to be dislodged by occasional errors, mutation, or drift."[263]

The paper also shows that "empathy allows populations to achieve high levels of cooperation even in the absence of a public [institution], and so there is less marginal benefit for empathetic individuals to

---

*prominence*, PHIL TRANS R SOC B 376 (2021) (discussing concept known as the Matthews Effect, where "low quality individuals are able to 'pass' as high quality based on their greater social prominence/capital, and a 'reputational poverty trap' where high quality individuals are unable to improve their social standing owing to a lack of social prominence/capital ")  There is also literature on the limits of empathy or outright even against empathy  *See generally* PAUL BLOOM, AGAINST EMPATHY: THE CASE FOR RATIONAL COMPASSION (2016) (making a non-partisan case against empathy and arguing that problems like climate change cannot be solved because of empathy); Molly Worthen, *The Trouble With Empathy*, N Y TIMES (Sept 4, 2020), https://www nytimes com/2020/09/04/opinion/sunday/empathy-school-college html (noting that it is impossible to inhabit another person's lived experience)  Both bodies of literature are beyond the scope of the Article, except to say that reputation and empathy are imperfect mechanisms, but compared to jail or fines, they are far better alternatives
[260] Radzvilavicius, Kessinger, & Plotkin, *supra* note 259
[261] *Id*. The authors note, however, "inferences about the perspective of another person are not always accurate, and the benefits of empathy are vulnerable to the possibility of deception and manipulation " *Id*.
[262] *Id*. at 4
[263] *Id*. (noting under a different model, Simple Standing, "strict institutions promote more cooperation because only those individuals who defect against good are labeled bad, and high values of *q* help keep these defectors in check ")

*GAME THEORY*

adhere to a public institution."[264] In the end, according to the paper, a society may voluntarily choose a system of "public monitoring" where members of the public pay a "tax" that is "shared equally among the Q members of the institution"—a calculus that even resists the possibility of "a bribe B provided $(1 – p) B < NT/Q$," an approach that "avoids the second-order free rider problem of costly punishment which would be required to explain the bottom-up formation of an honest institution in societies lacking structures for governance."[265] Thus, reputation and empathy hold the keys to designing large institutions that can effectively maintain cooperation when it has been established, whether or not there is some form of centralized authority.

There are examples of real-world institutions that use these mechanisms already. E-commerce companies "aggregate individual assessments of the reputations of buyers and sellers, providing a public broadcast to a large community of users."[266] While of course platforms like Amazon have faced problems with rankings being flooded with fake reviews,[267] companies like Google, Yelp, and ZocDoc have responded by verifying users to prevent this problem to increase trust in the marketplace, and claim to prevent companies from paying to have negative ratings removed. The Better Business Bureau acts as a non-profit watchdog agency for consumers and assigns scores to companies based on trustworthiness and will harm ratings based on meritorious disputes if businesses attempt to violate consumer protection laws.[268] Consumer Reports and Wirecutter offer professional evaluations of the quality of goods based on their independent third-party reviews.[269] Likewise, "credit bureaus synthesize and publicize the reputations of borrowers, so that lending agencies can choose to reward cooperative behavior with easy access to future capital.[270] Thus, reputation can be a powerful mechanism of earning trust, and trust can be rehabilitated even when it lapses through monitoring systems that teach users how to build their credit

---

[264] *Id.* at 7
[265] *Id.* at 9 *See also* Arunas L Radzvilavicius, Alexander J Stewart, & Joshua B Plotkin, *Evolution of Empathetic Moral Evaluation*, 8 EVOLUTIONARY BIOLOGY April 2019, at 1
[266] Radzvilavicius, Kessinger, & Plotkin, *supra* note 259, at 8
[267] Matt Stieb, *Amazon's War on Fake Reviews*, NEW YORK MAGAZINE (July 26, 2022), https://nymag com/intelligencer/2022/07/amazon-fake-reviews-can-they-be-stopped html
[268] BETTER BUSINESS BUREAU, https://www bbb org/search (last visited Jan 31, 2023)
[269] Consumer Reports, https://www consumerreports org (last visited Feb 1, 2023); Wirecutter, N Y TIMES, https://www nytimes com/wirecutter/ (last visited Feb 1, 2023)
[270] Radzvilavicius, Kessinger, & Plotkin, *supra* note 259, at 8

*GAME THEORY*

score effectively or allow both users and companies the opportunity to contest and correct bad reviews.

## CONCLUSION

To transform our legal profession, we must be willing to deeply engage with what it means to have "win-win" solutions in the context of lawyering. This Article does not purport to undertake such an exhaustive task. It concludes that although the Chicago School may have many flaws, ultimately, we engage in the same intellectual enterprise. In the words of Professor Eric Posner in his foundational book *Law and Social Norms*, the goal is "to lay bare the incentives that interfere with the perfect cooperation that would turn our flawed world into a utopian one" and that "[h]aving laid bare those incentives, one can proceed to investigate the mechanisms that enable cooperation where it would not otherwise exist."[271] The main difference is how we go about such an enterprise, and whether, in the end, we believe that is it possible to have institutions and a legal system based largely on cooperation, rather than on punishment.

The thesis of this paper is that game theory can be vital—when it comports with human intuition and scientific experiments. As economists like Professor Kaushik Basu have noted, the world needs a "rational rejection of rationality,"[272] i.e., a world where humans recognize it is to their mutual advantage to cooperate rather than defect, and where basic principles of justice, fairness, and equity are respected. As the research above shows, animals share these intuitions, and famous experiments demonstrate that monkeys treated differently than their neighbors will react negatively and engage in self harm because animals too have a basic sense of fairness.[273]

Groundbreaking work demonstrates that pure cooperation is possible, at least in math and nature. Mathematical models of human institutions also show that if we reexamine our assumptions of what it takes to maintain the fabric of society, we can form communities of repeat

---

[271] Posner, *supra* note 8, at 14-15 (also advocating for the use of reputation as an enforcement mechanism outside the government)

[272] BASU, *supra* note 42, at 244-262

[273] Megan van Wolkenten, Sarah F. Brosnan, & Frans B. M. de Waal, *Inequity responses of monkeys modified by effort*, 104 PROC. NAT'L ACAD. SCI. U S AM 18854 (PNAS) (2007) (famous experiment of monkeys treated differently than their neighbors by being given a less "valuable" reward—a cucumber instead of a grape—react negatively, and "only those pairs that spontaneously alternate high value rewards between themselves prove successful.")

## GAME THEORY

players with a "theory of the mind" that are self-aware. What might these "model communities" look like? The work of pioneering scholars such as Elinor Ostrom shows that they are all are around us.

What room does this leave for law and lawyering, and how does this instruct us to think about the larger project of reforming the tort and criminal law systems which are rationalized by the Chicago School using flawed math? One example of communities in need of reform is law schools. Many wise scholars currently recognize legal education as mis-education, where "any first-year law student . . . may begin her education imagining it as an invitation to ask fundamental questions concerning justice and power" but is likely to "'learn' quickly that serious legal thought in areas such as contracts and property prizes a certain vision of efficiency over all else" and that "contract law advances visions of equality that leave many forms of unequal power and vulnerability unchallenged or even enshrined as constitutionally fundamental."[274] Game theory can help us rethink the incentive structures on which legal education is based where even at the most basic level students are pitted against one and other for grades when education is based on a curve, instead of encouraged to work collaboratively in teams because what benefits one benefits all. And during the pandemic, much of legal education had to be rewritten, as classrooms went on Zoom, and curriculums were drastically adapted and threw traditional forms of instruction out the window.

This Article proposes that one day, using the cooperation insights found here, we can use new incentive structures and redesigned institutions to transform law schools from the alienating experience so many law students encounter into our own model communities: legal incubators and laboratories of change. It proposes that this is an advantageous moment to start thinking about the most valuable education we can provide our students so that we can offer it to them as so much of the existing curriculum is in desperate need of reform. The teachers I personally benefitted from the most did not teach me blackletter law on any given subject; they taught me how to deeply engage in a critical way with each and every single representation made as "truth"—they taught me to question everything and to strive for excellence and they taught me it was okay to fail and, rather than punish me for failing to conform to some ideal of perfection, they would reward the intellectual and emotional risks I undertook in giving my whole heart to my legal education.

---

[274] Purdy, Grewal, Kapczynski & Rahman, *supra* note 1, at 1827

*GAME THEORY*

It is the mentors who fought for me, and believed in me, even when I failed, that I strive to emulate.

As someone within the academy, I can only speak for what has worked for me in the classes I teach—a space where students know that just because I'm the teacher doesn't mean that I'm always right or that I'm perfect and incapable of mistakes. Rather, I draw on the pioneering work of Carol Dweck of Stanford and others in the field of educational research whose scholarship shows the deepest learning occurs through struggle.[275] I have had law students change my mind and introduce far wiser and smarter ideas than I've come up with and when they do, I applaud them because it seems to me learning has occurred.

If we in legal academia are ever to serve as our own model community where innovation and healing can take place, this must mean we listen, deeply listen to students. This is especially so for students from marginalized backgrounds who face real obstacles to being seen and witnessed in the classroom—like only having two gender-neutral bathrooms the entire building for students who are trans. Such an idea seems more reminiscent of *Hidden Figures*[276] but that's currently true of many institutions. I have so much idealism and hope—and it can bring us together when we start listening and treating our students as if they were capable of being *our* greatest teachers. I know many of the students I have worked with have been some of mine.

This "bottom up" law school—advocated by Critical Race Theorists like Gerald Lopez who teaches a class the way a conductor operates a symphony[277] and who is adamantly opposed to the Socratic Method[278]—would recognize part of the function of legal education is not teaching memorization, but creativity and innovation. These can occur where classrooms create safe spaces for all our students and staff. It means that the custodians who clean our buildings must be recognized for the valuable work they perform, as Duncan Kennedy suggested decades ago when he proposed they make the same salaries as law

---

[275] CAROL DWECK, MINDSET: THE NEW PSYCHOLOGY OF SUCCESS (2006)
[276] MARGOT LEE SHETTERLY, HIDDEN FIGURES: THE AMERICAN DREAM AND THE UNTOLD STORY OF THE BLACK WOMEN MATHEMATICIANS WHO HELPED WIN THE SPACE RACE 242-243 (2016)
[277] I am grateful to my colleague Steven Shiffrin for this observation
[278] Gerald P Lopez, *Transform—Don't Just Tinker With—Legal Education*, 23 CLINICAL L REV 471, 522 (2017) (critiquing the Socratic Method for promoting disengagement with the classroom); Gerald P Lopez, *Transform—Don't Just Tinker With—Legal Education (Part II)* 24 CLINICAL L REV 247 (2018); GERALD P LOPEZ, REBELLIOUS LAWYERING: ONE CHICANO'S VISION OF PROGRESSIVE LAW PRACTICE (1992)

*GAME THEORY*

professors[279]—that we must all learn each other's names and care about each other's wellbeing—and that we take care of one another when tragedy strikes, like raising funds for a colleague whose house burned down or providing support to teachers with young children during the pandemic. Such communities do exist within law schools—and they are often within the clinical programs. Recognizing the important work that clinicians do, the radical cutting-edge thought they teach, and the communication and self-reflection skills they foster—is vital to reawakening law school faculties where clinicians are often treated and seen as second-class citizens rather than the social justice warriors and the innovative teachers they are.

In light of these principles, it is clear that to be the communities we want to be, much of our existing legal education will need to be rethought and reimagined. An education may in many instances do more to shackle our minds than liberate them, so to start rethinking what we teach, we must also rethink who we want to be and what values and skills are most critical to impart. For a new generation of lawyers with artificial intelligence increasingly threatening to perform basic tasks lawyers carry out, we must teach them to be better—we must teach them, in addition to writing and research, self-reflection, self-care, emotional intelligence, creativity and innovation, and courage and boldness and perseverance and an ethic of community, and most importantly, that we are all in this together. For it is this generation of lawyers who may be faced with one of the most momentous tasks of history—a system redesign to prevent and to heal, rather than to punish, injustice.

---

[279] Duncan Kennedy, *Legal Education and the Reproduction of Hierarchy*, 32 J LEGAL EDUC 591 (1982)